\documentclass[twocolumn]{aastex63}

\usepackage{longtable}
\usepackage{graphicx}
\usepackage{amsmath,amssymb}
\usepackage{color}
\usepackage{units}
\usepackage{epstopdf}
\usepackage{hyperref}
\usepackage{multirow}

\newcommand{\beq}{\begin{equation}}
\newcommand{\eeq}{\end{equation}}
\newcommand{\bdm}{\begin{displaymath}}
\newcommand{\edm}{\end{displaymath}}
\newcommand{\T}[1]{\tilde{#1}}

\definecolor{Gray}{gray}{0.9}

\usepackage{dcolumn}
\newcolumntype{d}[1]{D{.}{.}{#1}}

\begin{document}

\title{The ZTF Source Classification Project: II. Periodicity and variability processing metrics}

\author[0000-0002-8262-2924]{Michael W. Coughlin}
\affiliation{School of Physics and Astronomy, University of Minnesota, Minneapolis, Minnesota 55455, USA}

\author{Kevin Burdge}
\affil{Division of Physics, Math, and Astronomy, California Institute of Technology, Pasadena, CA 91125, USA}

\author[0000-0001-5060-8733]{Dmitry A. Duev}
\affiliation{Division of Physics, Mathematics, and Astronomy, California Institute of Technology, Pasadena, CA 91125, USA}

\author{Michael L. Katz}
\affiliation{Department of Physics and Astronomy, Northwestern University, Evanston, IL 60208, USA}
\affiliation{Center for Interdisciplinary Exploration and Research in Astrophysics (CIERA), Evanston, IL 60208, USA}

\author[0000-0002-2626-2872]{Jan van Roestel}
\affil{Division of Physics, Math, and Astronomy, California Institute of Technology, Pasadena, CA 91125, USA}

\author{Andrew Drake}
\affil{Division of Physics, Math, and Astronomy, California Institute of Technology, Pasadena, CA 91125, USA}

\author[0000-0002-3168-0139]{Matthew J. Graham}
\affiliation{Division of Physics, Mathematics, and Astronomy, California Institute of Technology, Pasadena, CA 91125, USA}

\author{Lynne Hillenbrand}
\affil{Division of Physics, Math, and Astronomy, California Institute of Technology, Pasadena, CA 91125, USA}

\author{Ashish A. Mahabal}
\affil{Division of Physics, Math, and Astronomy, California Institute of Technology, Pasadena, CA 91125, USA}

\author[0000-0002-8532-9395]{Frank J. Masci}
\affiliation{IPAC, California Institute of Technology, 1200 E. California Blvd, Pasadena, CA 91125, USA}
             
\author[0000-0001-7016-1692]{Przemek Mr{\'o}z}
\affil{Division of Physics, Math, and Astronomy, California Institute of Technology, Pasadena, CA 91125, USA}

\author{Thomas A. Prince}
\affil{Division of Physics, Math, and Astronomy, California Institute of Technology, Pasadena, CA 91125, USA}

\author{Yuhan Yao}
\affil{Division of Physics, Math, and Astronomy, California Institute of Technology, Pasadena, CA 91125, USA}

\author[0000-0001-8018-5348]{Eric C. Bellm}
\affiliation{DIRAC Institute, Department of Astronomy, University of Washington, 3910 15th Avenue NE, Seattle, WA 98195, USA}

\author{Rick Burruss}
\affil{Caltech Optical Observatories, California Institute of Technology, Pasadena, CA 91125, USA}

\author{Richard Dekany}
\affil{Caltech Optical Observatories, California Institute of Technology, Pasadena, CA 91125, USA}

\author[0000-0002-3850-6651]{Amruta Jaodand}
\affiliation{Division of Physics, Math, and Astronomy, California Institute of Technology, Pasadena, CA 91125, USA}

\author[0000-0001-6295-2881]{David~L.\ Kaplan}
\affil{Department of Physics, University of Wisconsin-Milwaukee, Milwaukee, WI 53211, USA}

\author[0000-0002-6540-1484]{Thomas Kupfer}
\affiliation{Kavli Institute for Theoretical Physics, University of California, Santa Barbara, CA 93106, USA}

\author[0000-0003-2451-5482]{Russ R. Laher}
\affiliation{IPAC, California Institute of Technology, 1200 E. California
            Blvd, Pasadena, CA 91125, USA}

\author{Reed Riddle}
\affil{Caltech Optical Observatories, California Institute of Technology, Pasadena, CA 91125, USA}

\author[0000-0002-8121-2560]{Mickael Rigault}
\affil{Universit\'e Clermont Auvergne, CNRS/IN2P3, Laboratoire de Physique de Clermont, F-63000 Clermont-Ferrand, France}

\author{Hector Rodriguez}
\affil{Caltech Optical Observatories, California Institute of Technology, Pasadena, CA 91125, USA}

\author[0000-0001-7648-4142]{Ben Rusholme}
\affiliation{IPAC, California Institute of Technology, 1200 E. California
             Blvd, Pasadena, CA 91125, USA}

\author{Jeffry Zolkower}
\affil{Caltech Optical Observatories, California Institute of Technology, Pasadena, CA 91125, USA}

\begin{abstract}
The current generation of all-sky surveys is rapidly expanding our ability to study variable and transient sources.
These surveys, with a variety of sensitivities, cadences, and fields of view, probe many ranges of timescale and magnitude.
Data from the Zwicky Transient Facility (ZTF) yields an opportunity to find variables on timescales from minutes to months. 
In this paper, we present the codebase, \texttt{ztfperiodic}, and the computational metrics employed for the catalogue based on ZTF's Second Data Release.
We describe the publicly available, graphical-process-unit optimized period-finding algorithms employed, and highlight the benefit of existing and future graphical-process-unit clusters.
We show how generating metrics as input to catalogues of this scale is possible for future ZTF data releases.
Further work will be needed for future data from the Vera C. Rubin Observatory's Legacy Survey of Space and Time.
\end{abstract}


\section{Introduction}

The study of variable and transient sources is rapidly expanding based on the large data sets available from wide-field survey telescopes. Amongst others these include the Panoramic Survey Telescope and Rapid Response System (Pan-STARRS; \citealt{MoKa2012}), the Asteroid Terrestrial-impact Last Alert System (ATLAS; \citealt{ToDe2018}), the Catalina Real-Time Transient Survey (CRTS; \citealt{DrDj2009,DrGr2014}), the All-Sky Automated Survey for SuperNovae (ASAS-SN; \citealt{ShPr2014,KoSh2017}), the Zwicky Transient Facility (ZTF; \citealt{Bellm2018,Graham2018,DeSm2018,MaLa2018}), the Visible and Infrared Survey Telescope for Astronomy (VISTA; \citealt{LoCr2020}) and in the near future, the Vera C. Rubin Observatory's Legacy Survey of Space and Time (LSST; \citealt{Ivezic2014}). Catalogs of variable sources from these surveys are building upon the highly successful work identifying periodic variable stars in the Magellanic Clouds and
Galactic center; we highlight catalogs such as those from MACHO \citep{AlAl2000} and OGLE \citep{udalski2003,udalski2015} as examples.
Coupled with Gaia's second data release (DR2; \citealt{Gaia2018}) and soon DR3, we are currently experiencing a dramatic expansion of our capabilities for doing time-domain astrophysics.
The range of sensitivities and cadences change the parameter space of magnitude ranges and timescales probed, which can span from years down to minutes for many of these surveys.

\begin{figure*}[t]
\centering
\includegraphics[width=6.0in]{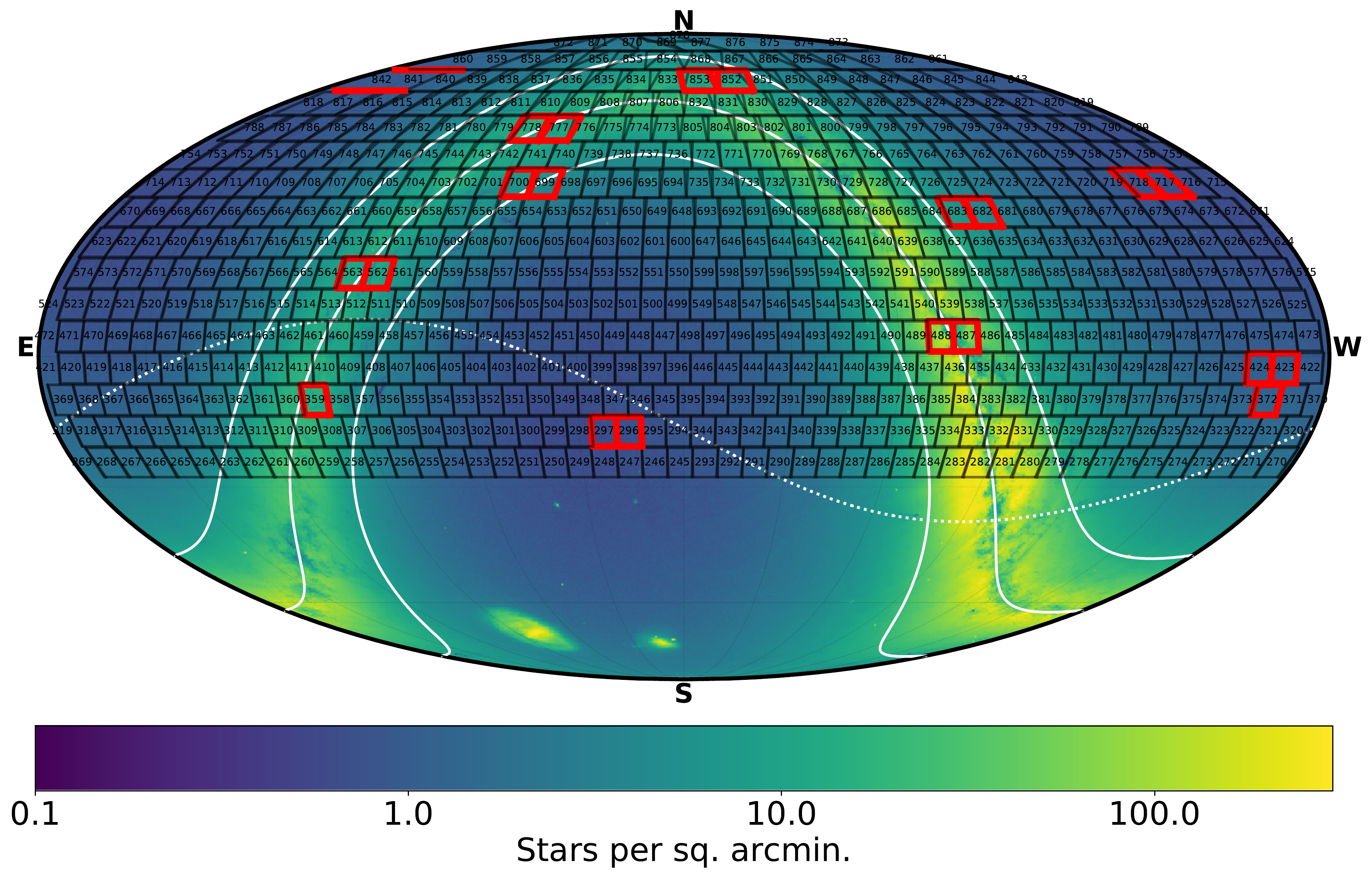}
\caption{Mollweide projection in equatorial coordinates of the ZTF field coverage as a function of stellar density, using Gaia DR2 \protect\citep{Gaia2018}. The black lines correspond to ZTF field boundaries, with the number contained corresponding to the field number. The solid white lines correspond to galactic latitude of $-$15, 0, and 15, while the dashed line corresponds to the ecliptic. We highlight the 20 fields in the initial catalog release in red (van Roestel et al. in prep).}
\label{fig:coverage}
\end{figure*}

While these surveys can be used to follow-up objects of interest, here, we are primarily interested in ``untargeted'' searches, i.e., searches without a priori knowledge about objects, for variable and periodic objects.
These searches enable identification of populations of objects, such as cataclysmic variables \citep{SzDi2020}, Be stars \citep{NgLe2019}, Cepheids and RR Lyrae, useful for studying the processes and properties of galaxy formation \citep{Sa1984,Sa1985,Ca2009} and measuring the expansion rate of the Universe \citep{FrMa2001}.
In addition, they are also sensitive to rare objects, such as short-period white dwarf binaries \citep{BuCo2019,BuFu2019,CoBu2020}, large-amplitude, radial-mode hot subdwarf pulsators \citep{KuBa2019}, and ultracompact hot subdwarf binaries \citep{KuBa2020}.

To highlight one source class, we are discovering and characterizing the population of so-called ultra-compact binaries (UCBs), which have two stellar-mass compact objects with orbital periods $P_{\mathrm{orb}}$ $<$\,1 hour \citep{BuCo2019,BuFu2019,CoBu2020}.
Many of the UCBs emit gravitational waves in the milliHertz regime with sufficient strain for the upcoming \emph{Laser Interferometer Space Antenna} (\emph{LISA}) to detect \citep{AmAu2017}. These \emph{LISA} ``verification sources'' will serve as crucial steady-state sources of gravitational waves that will not only verify that \emph{LISA} is operating as expected \citep{KuKo2018}, but also themselves serve as probes of binary stellar evolution \citep{NeTo2005,KrCh2018,Ban2018,AnTo2017}, white dwarf structure \citep{FuLa2011}, Galactic structure \citep{BrMi2019}, accretion physics \citep{CaNe2015} and general relativity \citep{BuCo2019,KuBa2019}.

We are interested in computationally efficient algorithms for measuring the variability and periodicity of large suites of light curves for the purpose of creating catalogs of sources.
These catalogs are useful for identifying classes of sources such as those mentioned above, as well as for mitigating the presence of known variable sources in searches for new transient objects, such as short $\gamma$-ray burst or gravitational-wave counterparts, e.g., \citet{CoAh2019,CoAh2019b}. 
The size of all-sky survey data are large, enabling identification of a large number of sources. However, the computational cost scales linearly with the number of objects.
In ZTF's Second Data Release (DR2)\footnote{\url{https://www.ztf.caltech.edu/page/dr2}}, which we will employ here, there are 3,153,256,663 light curves, regardless of passband; for comparison, a cross-match to a Pan-STARRS catalog indicates that $\sim$\,1.3\,billion of these sources are unique.
In the following, each individual light curve will be processed if they exceed 50 detections after surviving data quality and other cuts described below; we will focus our discussion in this paper on a small fraction of the fields, covering a range of stellar densities, consistent with those chosen for classification studies in the accompanying catalog paper (van Roestel et al. in prep).
We note that this separately treats $g$-, $r$-, and $i$-band observations of the same light curves, and the set includes many objects with hundreds of detections in multiple passbands.
In the following, we analyze single-band lightcurves, and so the same objects in different passbands are analyzed separately.
We seek to choose variability metrics and periodicity algorithms appropriate for analyzing this data-set.
We have taken inspiration from variability and periodicity codebases, such as \texttt{FATS} \citep{NuPr2015}, \texttt{astrobase} \citep{BhBo2020} and \texttt{cesium} \citep{NaWa2016} in choosing the metrics, with a focus on using scalable graphical processing unit (GPU) periodicity algorithms to make the required large-scale processing tractable.

In this paper, we will describe the pipeline \texttt{ztfperiodic}\footnote{\url{https://github.com/mcoughlin/ztfperiodic}}, which we use to systematically identify variable and periodic objects in ZTF; we use a mix of metrics designed to efficiently identify and characterize variable objects as well as algorithms to phase-fold light-curves in all available passbands. We use a variety of available arrays of GPUs to period search all available photometry. The arrays of GPUs available are ideal for period finding large data sets and already have shown significant speed-ups relative to central processing units (CPUs).

\section{Observational Data}

\begin{table}[]
\hspace*{-1.7cm}
\begin{tabular}{c d{3.2} d{3.2} d{3.2} d{3.2}}
\hline
Field & \multicolumn{1}{c}{RA [$^\circ$]} & \multicolumn{1}{c}{Dec [$^\circ$]} & \multicolumn{1}{c}{Gal. Long. [$^\circ$]} & \multicolumn{1}{c}{Gal. Lat. [$^\circ$]} \\ \hline
296   & 15.79  & -17.05 & 141.27 & -79.14  \\ \hline
487   & 281.19 & 4.55 & 36.55 & 3.03 \\ \hline
682   & 266.86 & 33.35  & 58.61 & 26.73    \\ \hline
851   & 351.43 & 69.35  & 115.73 & 7.93  \\ \hline
\end{tabular}
\caption{Locations of the fields (measured at their centers) highlighted in this paper.}
\label{tab:fields}
\end{table}

\begin{figure*}[t]
\includegraphics[width=3.5in]{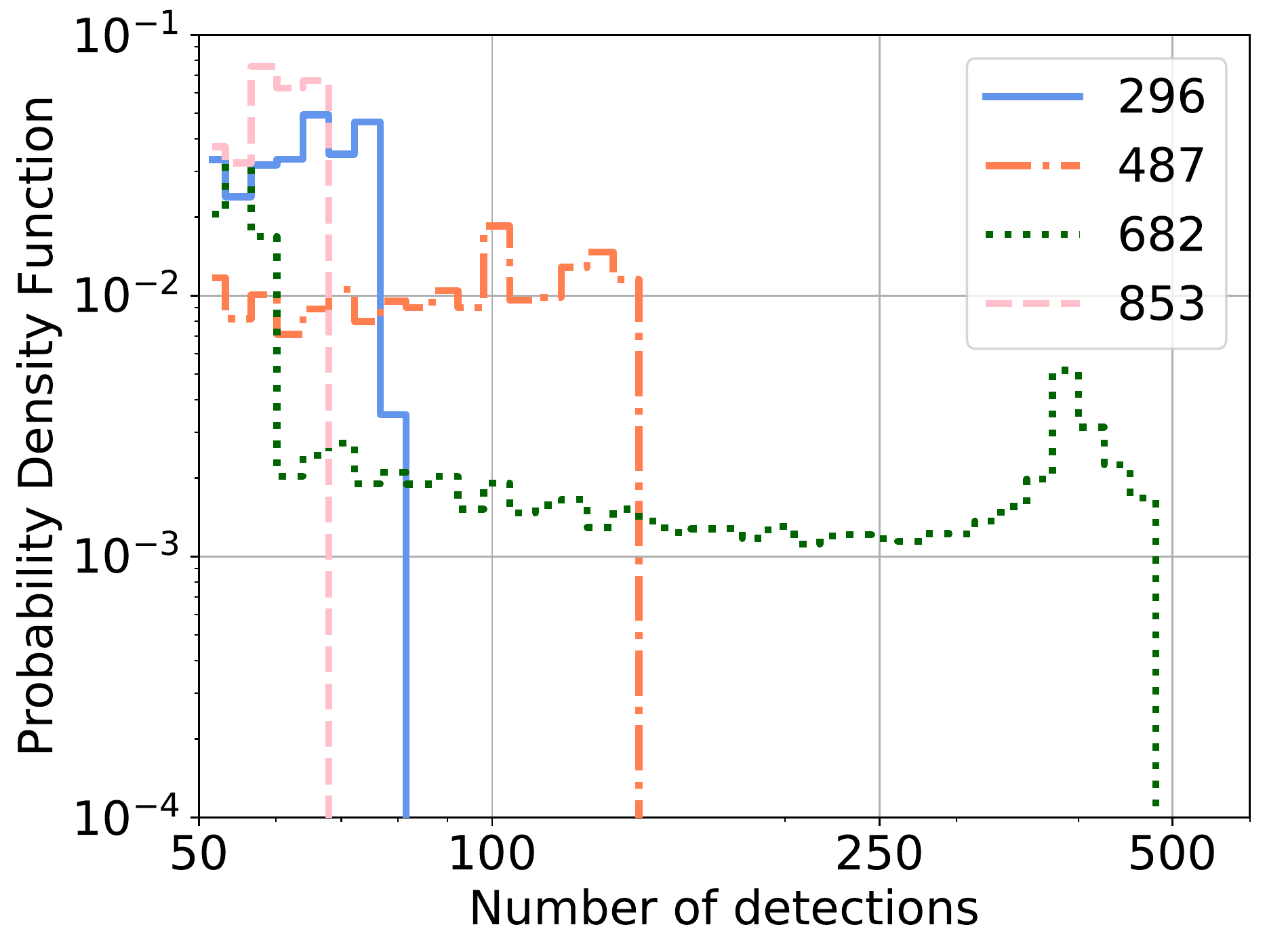}
\includegraphics[width=3.5in]{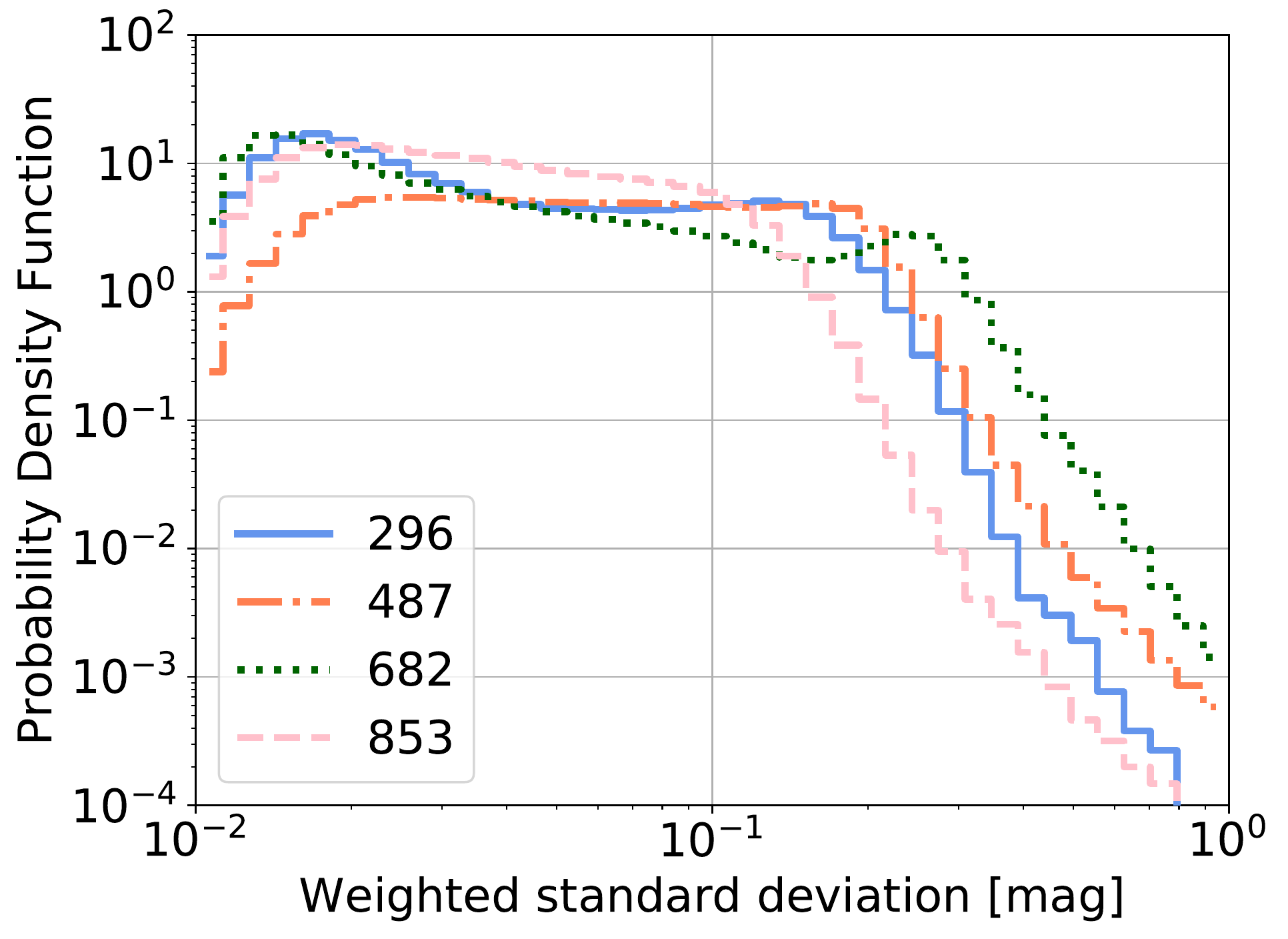}
\caption{Left: Probability density function of the number of detections for individual objects passing the data quality and time cuts for four example fields. The field IDs are given in the legend, and their locations given in Table~\ref{tab:fields}. Right: Probability density function of the weighted standard deviations for the light curves of individual objects.}
\label{fig:Nobs}
\end{figure*}

We employ data predominantly from DR2 in this analysis, which covers public data between 2018-03-17 and 2019-06-30, and private data between 2018-03-17 to 2018-06-30; in our analysis, we also include private data up until 2019-06-30.
ZTF's Mid-Scale Innovations Program in Astronomical Sciences (MSIP) program covers the observable night sky from Palomar with a 3 night cadence, and a nightly cadence in both $g$ and $r$ filters in the Galactic plane \citep{BeKu2019}. 
The ZTF partnership is also conducting a moderate cadence survey over a 3000 square degree field at high declination, visiting the entire field 6 times per night in $g$ and $r$ filters, resulting in more than 1000 epochs.
Within this field, we expect to probe variables at the 20 mmag level for objects at 16th magnitude, and around 100 mmag at 20th magnitude \citep{MaLa2018}.
Also, ZTF conducts a high cadence survey in the Galactic plane, with galactic latitude $|b|<14^{\circ}$ and galactic longitude $10^{\circ}<l<230^{\circ}$, where the camera continuously sits on a field for several hours with a 40\,s cadence.
Overlayed by stellar densities, we show the ZTF primary field grid in Figure~\ref{fig:coverage}; we note there is also a secondary grid filling in gaps in the primary grid. In this Figure, we also include outlined in red the fields targeted in the forthcoming catalog paper (van Roestel et al. in prep).

The left panel of Figure~\ref{fig:Nobs} shows the probability density function of the number of detections for individual objects passing the data quality and time cuts for four example fields. 
The locations of these fields are given in Table~\ref{tab:fields}, chosen to span a variety of galactic latitudes and longitudes and therefore stellar densities. 
This analysis removes any observation epochs indicating suspect photometry according to the {\it catflags} value in the light curve metadata\footnote{For details, see the DR2 Release Summary \url{https://www.ztf.caltech.edu/page/dr2}, specifically Section 9b.}.
It also removes any ``high cadence'' observations, defined as observations within 30\,minutes of one another; for any series of observations with observation times less than 30 minutes apart, we keep only the first observation in the series.
This is to support the periodicity analyses; an over-abundance of observations in very densely observed sets otherwise dominate the period finding statistics, as these observations will have the same weight as all temporally separated observations, but lack the sensitivity to periods longer than the single night in which they are taken.
This short-coming will be a point of study moving forward.
In addition to those observations removed, some subset of sources have fewer observations within the same field; these occur because faint stars have fewer detections because of lower detection efficiency. 
We also remove any stars within 13$^{\prime\prime}$ of either an entry in the Yale Bright Star Catalog \citep{HoJa1991} and Gaia  \citep{Gaia2018} stars brighter than 13th magnitude; we arrived at this value through evaluating the measured variability of stars near cataloged objects, and $13^{\prime\prime}$ was a threshold beyond which the variability was consistent with background.
This helps to remove bright blends from the variability selection process.
On the right of Figure~\ref{fig:Nobs}, we show the probability density function of the weighted standard deviation for the light curves of individual objects in the example fields, showing consistency of this metric across fields.

\begin{figure}[t]
\includegraphics[width=3.5in]{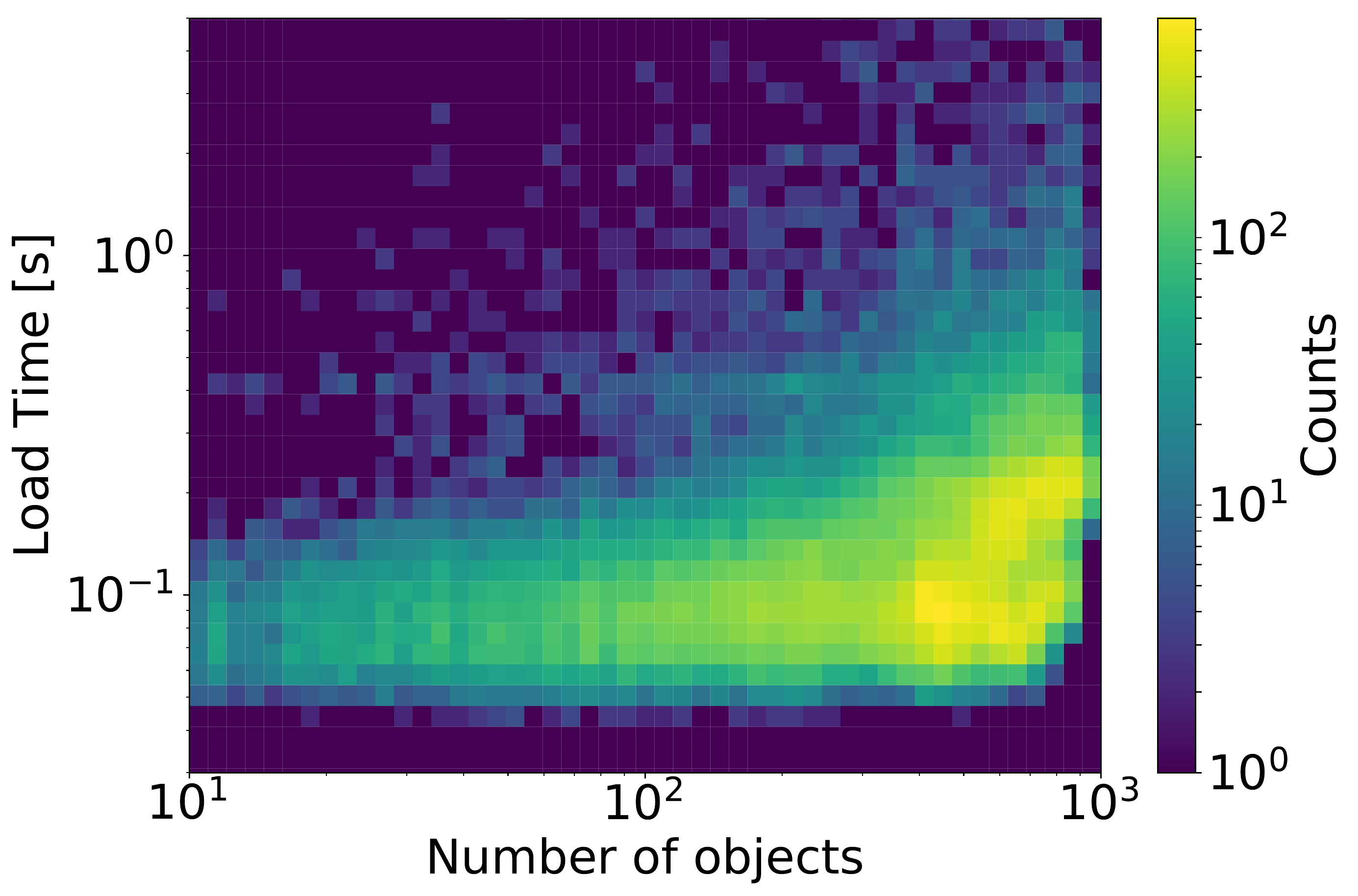}
\caption{Two dimensional histogram of the read time from the ZTF light curve database ``Kowalski'' vs. the number of objects returned. The load time is mostly independent of the number of objects returned.}
\label{fig:kowalski}
\end{figure}

To query the photometry, we are using ``Kowalski''\footnote{https://github.com/dmitryduev/kowalski}, an efficient non-relational database that uses MongoDB to efficiently store and access both ZTF alert/light curve data and external catalogs (including 230M+ alerts and 3.1B+ light curves) \citep{DuMa2019}.
We show the read time from the ZTF light curve database ``Kowalski'' vs. the number of objects returned on the left of Figure~\ref{fig:kowalski}.
We note that the typical 100-200\,ms includes the typical 54\,ms latency between California (where Kowalski resides) and Minnesota (where the test was performed), including $\sim$\,20\,ms of light travel time.
As can be seen, the analysis takes $\sim$\,1\,s per light curve across the range of light curves analyzed at a single time (100 to 1000 which is the memory limit for the GPUs).

The light curves are analyzed in groups of $\sim$\,1000 to fill out the RAM available on most GPUs employed; when the jobs are being distributed, the total number of light curves in a given quadrant on a particular CCD within a field are queried. This is possible as ZTF has two grids on which all observations are performed, with a repeating pointing accuracy varying $\sim$\,50-200$^{\prime\prime}$; therefore, a particular object will generally only appear twice, once on each grid. For reference, there are 64 quadrants, 4 quadrants for each of the 16 CCDs. Based on this, the number of jobs required to analyze the total number of light curves, in chunks of 1000, are computed. To ensure the efficacy of the variability and periodicity metrics we discuss below, we place a threshold of at least 50 detections in a single band for an object to be analyzed. This requirement means fewer than 1000 light curves are sometimes analyzed because some will not meet the 50 observation limit required for analysis. Because \texttt{ztfperiodic} analyzes the data in chunks like these, it is simple to parallelize across GPUs, with different GPUs running on different sets of $\sim$\,1000 light curves; an hdf5 file is written at the end to disk containing the statistics for a particular job (see below), each chunk receiving a different name based on a simple convention (field, CCD, quadrant, job index).

\section{Variable Source Metrics}

Due to the significant amount of astronomical time series data provided by ZTF and other all-sky surveys, it is important to have robust selection criteria and algorithms to find variable objects.
In ground-based surveys like ZTF, light curves are irregularly sampled, have gaps, and can have large statistical errors, which means that these algorithms must be robust in order to efficiently find true signals.
The catalog includes two main types of metrics: variability and periodicity.
These metrics are typically useful for machine learning algorithms to group light curves into categories through feature extraction based on light curve data.
The goal of these features is to encode numerical or categorical properties to characterize and distinguish the different variability classes, such that machine learning algorithms can distinguish between classes of light curves.

There are a variety of computationally cheap variability metrics we use, ranging from relatively basic statistical properties such as the mean or the standard deviation to more complex metrics such as Stetson indices \citep{St1996}. As shown by \cite{PaSo2017}, many of the commonly used features are strongly correlated. We therefore chose the set of features suggested by \cite{PaSo2017}, with the addition of robust measurements of the amplitude. We also performed a Fourier-decomposition to characterize the shape of the folded light curve better.
We detail the choices we have made below, which include the number of measurements, weighted mean and median magnitudes (RMS and percentile-based), kurtosis, skewness, variance, chi-square, Fourier indices, amongst many others.
The statistics simply require three vectors for each light curve, encoding the time, magnitude, and magnitude error.

\begin{figure}[t]
\includegraphics[width=3.5in]{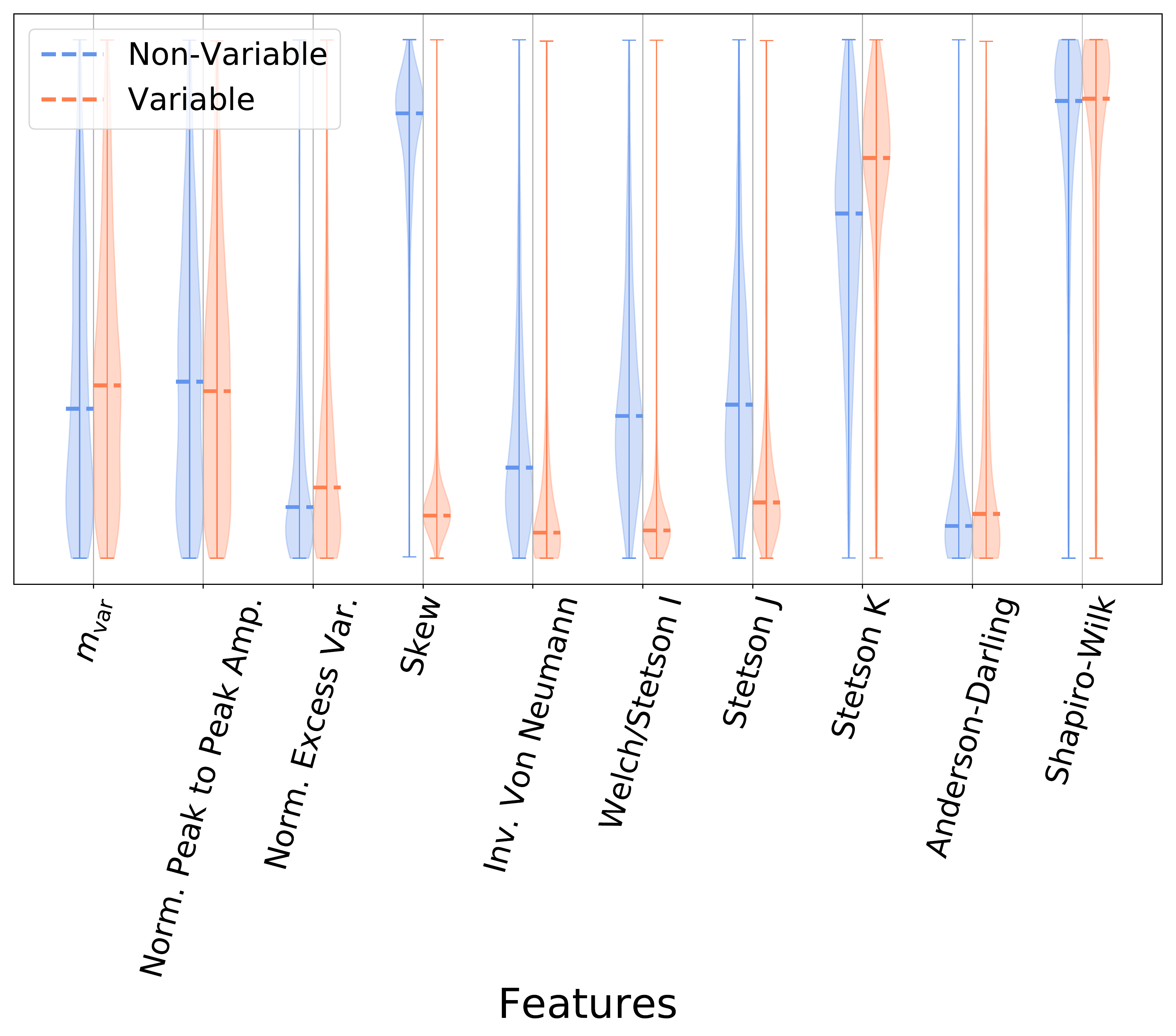}
\caption{Probability density for a subset of the features in the analysis for a ``variable'' and ``non-variable'' set of objects (van Roestel et al. in prep), as assessed by a machine learning algorithm XGBoost \protect\citep{ChGu2016}. The dashed lines correspond to the 50th percentile for the features. For simplicity, we have normalized all features such that they appear on the same plot, with values increasing in the upwards direction on the otherwise arbitrary y-axis.}
\label{fig:features}
\end{figure}

We provide a summary of the metrics employed in Table~\ref{table:statistics}. While the accompanying catalog paper (van Roestel et al. in prep) will discuss their use extensively, to demonstrate their efficacy, Figure~\ref{fig:features} shows the probability density for a subset of the features in the analysis for a ``variable'' and ``non-variable'' set of objects. We define ``variable'' objects as those that statistically change from observation to observation inconsistent with their assigned error bars; ``non-variable'' objects do not display these traits at the signal-to-noise reached by ZTF.
A number show clear differentiation between the two sets, indicating their suitability as metrics for this purpose. We note that while some metrics have similar marginalized distributions, this marginalized format masks potential higher order correlations between parameters, and therefore we choose to use all metrics in our classifier. 

\begin{table*}[]
\centering
\begin{tabular}{|c|c|c|}
\hline
Index & Statistic                                             & Calculation                                                                                                                                                                                                                                                                                                                    \\ \hline
1     & $N$                                                   & Number of observations passing cuts                                                                                                                                                                                                                                            \\ \hline
2     & $m_{\textrm{median}}$                                 & Median magnitude                                                                                                                                                                                                                                                                                                               \\ \hline
3     & $m_{\textrm{mean}}$                                   & Weighted mean                                                                                                                                                                                                                                                                                                                  \\ \hline
4     & $m_{\textrm{var}}$                                    & Weighted variance                                                                                                                                                                                                                                                                                                              \\ \hline
5     & $\chi^2$                                              & $\frac{1}{N-1} \sum\limits_{i} \frac{\left(m_{\textrm{median}}-m_i \right)^{2}}{\sigma_i^2}$                                                                                                                                                                                                                                   \\ \hline
6     & RoMS                                                  & $\frac{1}{N-1} \sum\limits_{i} \frac{\left|m_{\textrm{median}}-m_i \right|}{\sigma_i}$                                                                                                                                                                                                                                         \\ \hline
7     & Median absolute deviation                             & median $\left( |m - m_{\textrm{median}}| \right)$                                                                                                                                                                                                                                                                              \\ \hline
8     & Normalized Peak to Peak Amplitude                     & $\frac{\max(m-\sigma) - \min(m+\sigma)}{\max(m-\sigma) + \min(m+\sigma)}$                                                                                                                                                                                                                                                      \\ \hline
9     & Normalized Excess Variance                            & $\frac{1}{N m_{\textrm{mean}}^2} \sum\limits_{i} \left(m_{\textrm{mean}}-m_i \right)^{2} - \sigma_i^2$                                                                                                                                                                                                                         \\ \hline
10-14 & Ranges                                                & Inner 50\%, 60\%, 70\%, 80\%, and 90\% Range                                                                                                                                                                                                                                                                                   \\ \hline
15    & Skew                                                  & $\frac{N}{(N-1) (N-2)} \sum\limits_{i} \frac{\left(m_{\textrm{mean}}-m_i \right)^{3}}{\sigma_i^3}$                                                                                                                                                                                                                             \\ \hline
16    & Kurtosis                                              & $\frac{N (N+1)}{(N-1) (N-2) (N-3)} \sum\limits_{i} \frac{\left(m_{\textrm{mean}}-m_i \right)^{4}}{\sigma_i^4} - \frac{3 \left(N-1\right)^{2}}{(N-2) (N-3)}$                                                                                                                                                                    \\ \hline
17    & Inverse Von Neumann Statistic                         & $\eta = \left( \frac{1}{\sum\limits_{i} \left(\frac{1}{\Delta t_i} \right)^2 m_{\textrm{var}}} \right) \sum\limits_{i} \left( \frac{\Delta m_i}{\Delta t_i} \right)^2$ where $\Delta t_i = t_{i+1} - t_{i}$ and $\Delta m_i = m_{i+1} - m_{i}$                                                                                 \\ \hline
18    & Welch/Stetson I                                       & $\frac{N}{N-1} \sum\limits_{i} \left( \frac{m_i - m_{\textrm{mean}}}{\sigma_i} \right) \left( \frac{m_{N-i} - m_{\textrm{mean}}}{\sigma_{N-i}} \right)$                                                                                                                                                                        \\ \hline
19    & Stetson J                                  & $\sqrt{\frac{N}{N-1}} \sum\limits_{i} sgn \left( (m_i - m_{\textrm{mean}}) (m_{N-i} - m_{\textrm{mean}}) \right) \sqrt{\left| \left( \frac{m_i - m_{\textrm{mean}}}{\sigma_i} \right) \left( \frac{m_{N-i} - m_{\textrm{mean}}}{\sigma_{N-i}} \right) \right|}$                                                                \\ \hline
20    & Stetson K                                  & $\sqrt{\frac{1}{N}} \sum\limits_{i} \left| \left( \frac{m_i - m_{\textrm{mean}}}{\sigma_i} \right) \left( \frac{m_{N-i} - m_{\textrm{mean}}}{\sigma_{N-i}} \right) \right| / \sqrt{\sum\limits_{i} \left( \frac{m_i - m_{\textrm{mean}}}{\sigma_i} \right) \left( \frac{m_{N-i} - m_{\textrm{mean}}}{\sigma_{N-i}} \right)^2}$ \\ \hline
21    & Anderson-Darling test &       \cite{St1974}                                                                                                                                                                                                                                                                                                                         \\ \hline
22    & Shapiro-Wilk test   &       \cite{ShWi1965}                                                                                                                                                                                                                                                                                                                         \\ \hline
23-35 & Fourier Decomposition                                 & $y = c_1 + c_2 \times t + \sum\limits_{i=1}^{5} \left(a_i \cos(\frac{2 \pi t i}{P}) + b_i \sin(\frac{2 \pi t i}{P}) \right)$                                                                                                                                                                                                                                   \\ \hline
36    & Bayesian Information Criterion                        & $\chi^2$ likelihood keeping different number of Fourier harmonics                                                                                                                                                                                                                                                              \\ \hline
37    & Relative $\chi^2$                                     & $y = c_1 + c_2 \times t$ to full Fourier Decomposition                                                                                                                                                                                                                                                                             \\ \hline
\end{tabular}
\caption{Statistics calculated based on the light curves and period finding. $N$ is the number of observations, $m_i$ is the $i$th observation magnitude, and $\sigma_i$ is the $i$th observation magnitude error. The variables $c_1$, $c_2$ and the $a_i$'s and $b_i$'s are constants to be fit for in statistics 23-35 and 37. In addition to these statistics, we save the best fit period based on the hierarchical analysis as well as its significance.}
\label{table:statistics}
\end{table*}

\section{Period Finding}

\begin{figure*}[t]
\includegraphics[width=3.5in]{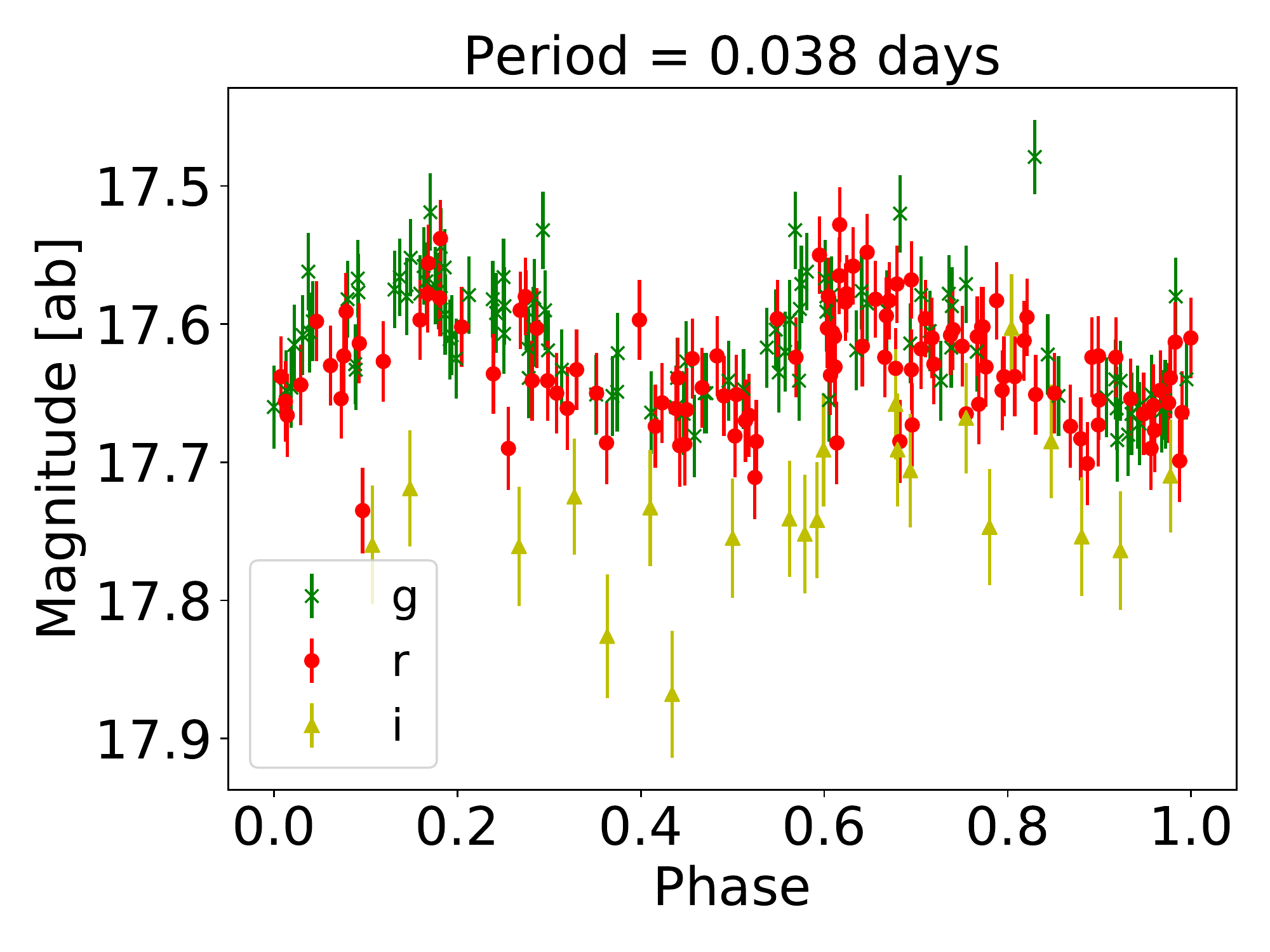}
\includegraphics[width=3.5in]{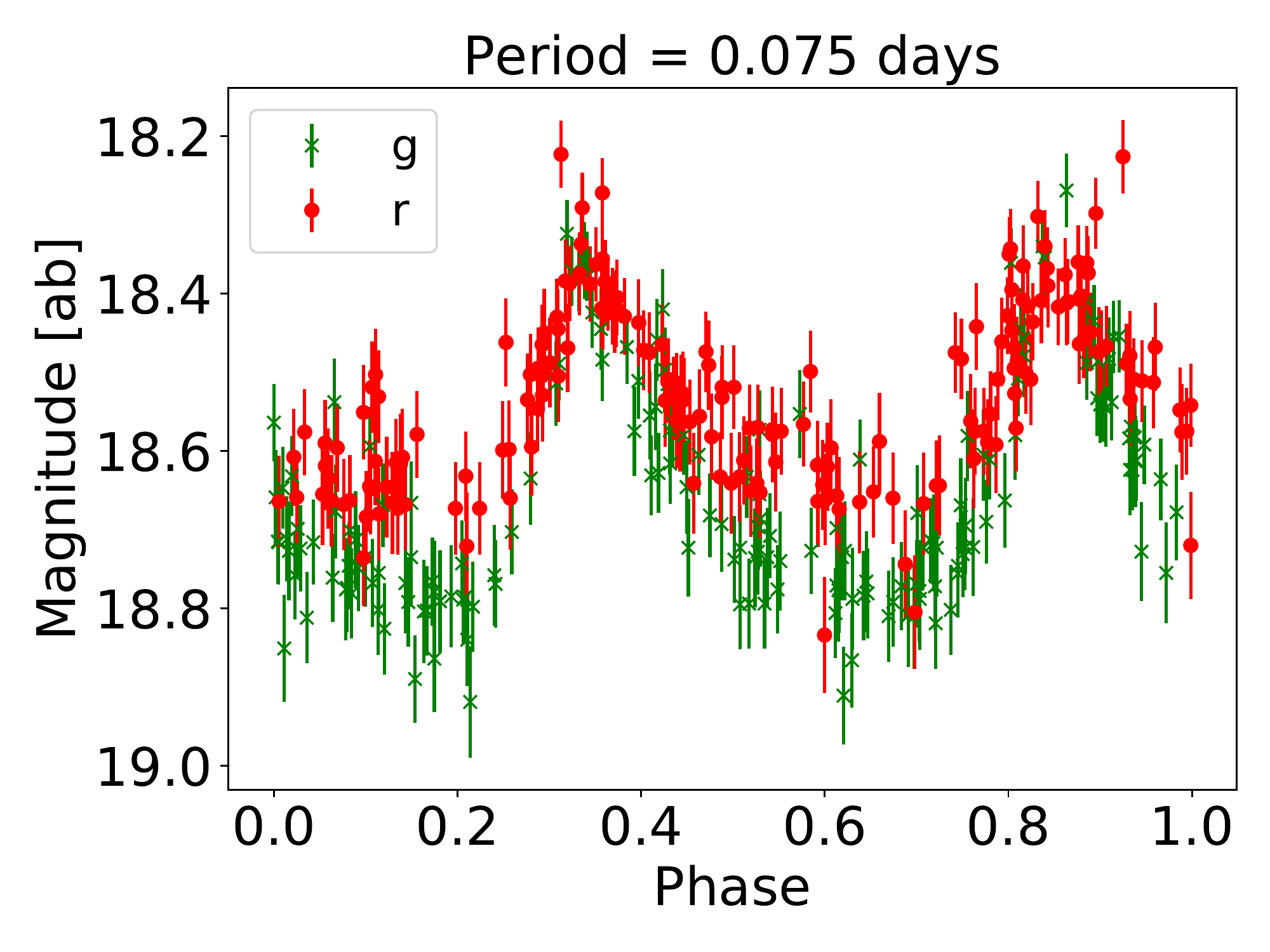}
\includegraphics[width=3.5in]{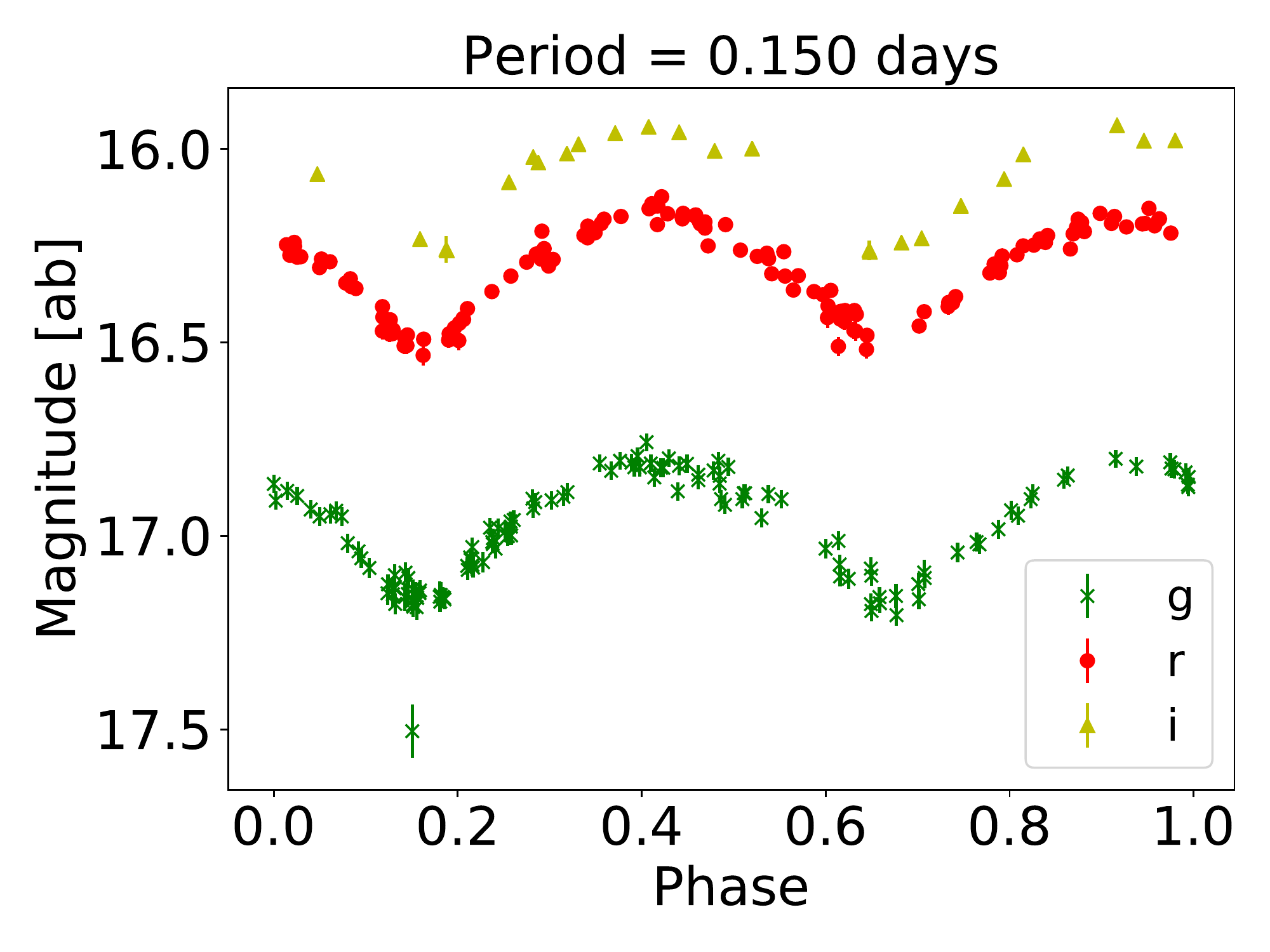}
\includegraphics[width=3.5in]{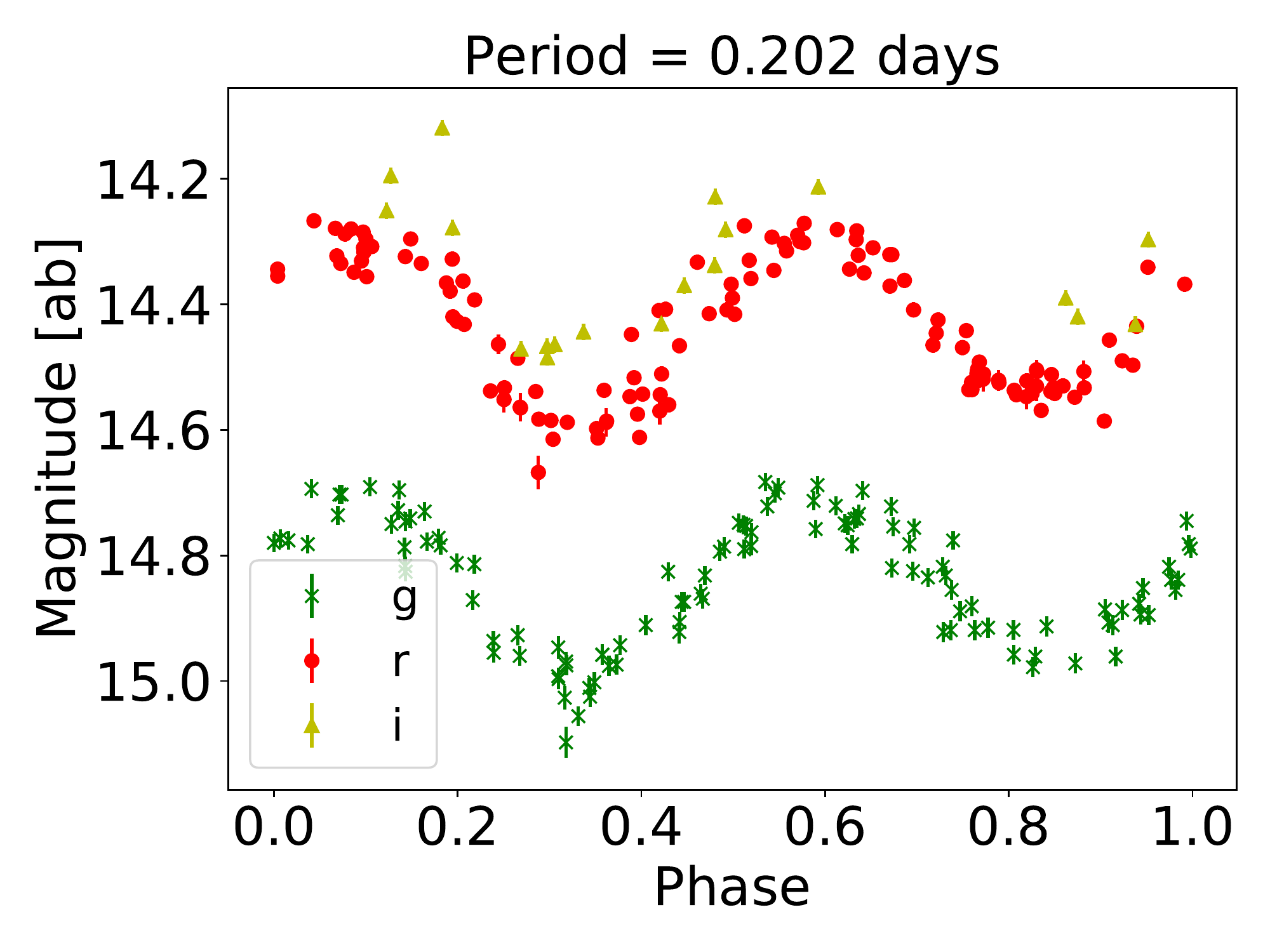}
\includegraphics[width=3.5in]{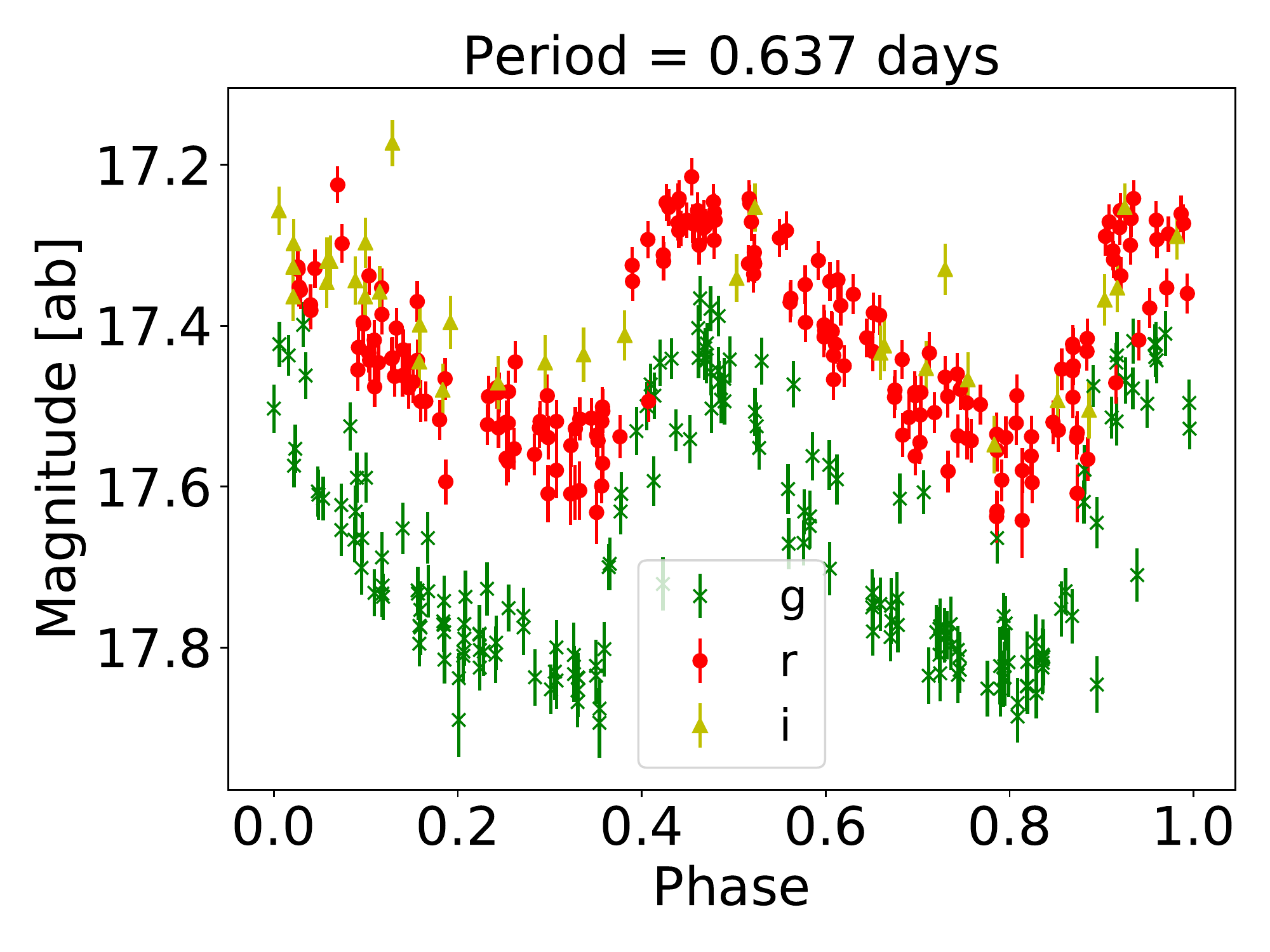}
\includegraphics[width=3.5in]{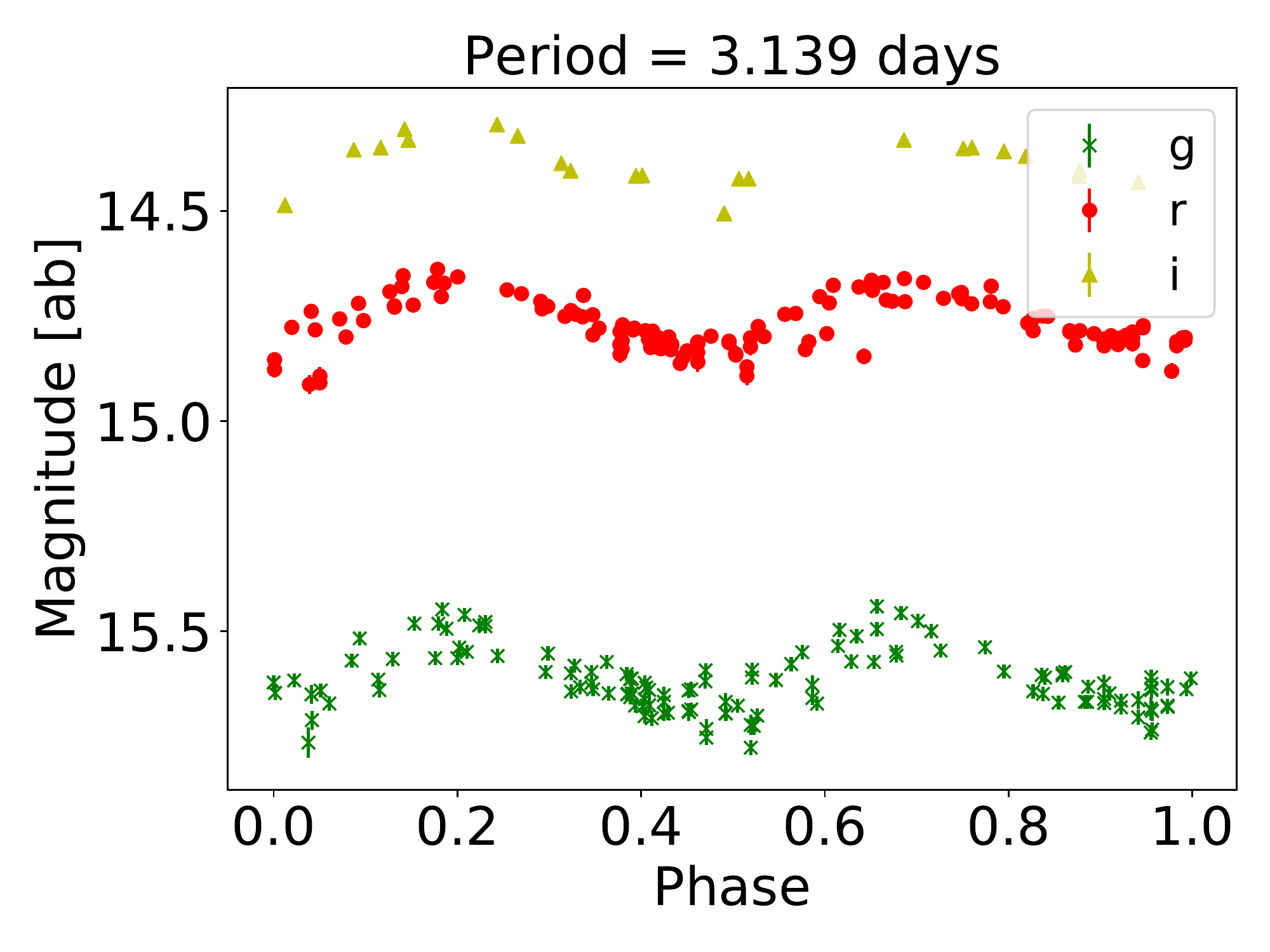}
\caption{Example periodic variables (their computed periods are given in the titles) folded at twice their computed period, identified in the various bins we use for scanning. We fold at twice the computed period to evaluate the consistency of the reconstructed light curves across two orbits, useful for assessing the correctness of the period. These bins are chosen to simplify comparison of periodicity significances estimated between source variable on very different time scales: top left, 30--60\,min (delta Scuti), top right, 1--2\,hr (delta Scuti), middle left 2--4\,hr (W UMa), middle right, 4--12\,hr (W UMa), bottom left, 12--72\,hr (RRab star) and bottom right, 3--10\,days (Cepheid). For W UMa's, the true period is two times longer than the computed period.}
\label{fig:examples}
\end{figure*}

We use period-finding algorithms to estimate periods for all objects; phase-folded light-curves, such as those in Figure~\ref{fig:examples}, are used when scanning and classifying sources, typically for those with significant periodicities.
There are a variety of period finding algorithms in the literature (see \citealt{GrDr2013b} for a comparison and review), including those based on least-squares fitting to a set of basis functions \citep{ZeKu2009,MoFa2015,MoCo2017}.
By far the dominant source of computational burden is in the period finding. The algorithm we employ is hierarchical, with two period finding algorithms supplying candidate frequencies to a third. 

The first algorithm is a conditional entropy (CE; \citealt{GrDr2013}) algorithm.
CE is based on an information theoretic approach; broadly, information theory-based approaches improve on other techniques by capturing higher order statistical moments in the data, which are able to better model the underlying process and are more robust to noise and outliers.
Graphically, CE envisions a folded light curve on a unit square, with the expectation that the true period will result in the most ordered arrangement of data points in the region.
More specifically, the algorithm minimizes the entropy associated with the system relative to the phase, which naturally accounts for the non-trivial phase space coverage of real data.
The CE version we use is \texttt{gce} \citep{KaCo2020}\footnote{https://github.com/mikekatz04/gce}. It is implemented on Graphics Processing Units (GPU) in CUDA \citep{CUDA} and wrapped to \texttt{Python} using Cython \citep{Cython} and a special CUDA wrapping code \citep{CUDAwrapper}.

\begin{figure*}[t]
\includegraphics[width=3.5in]{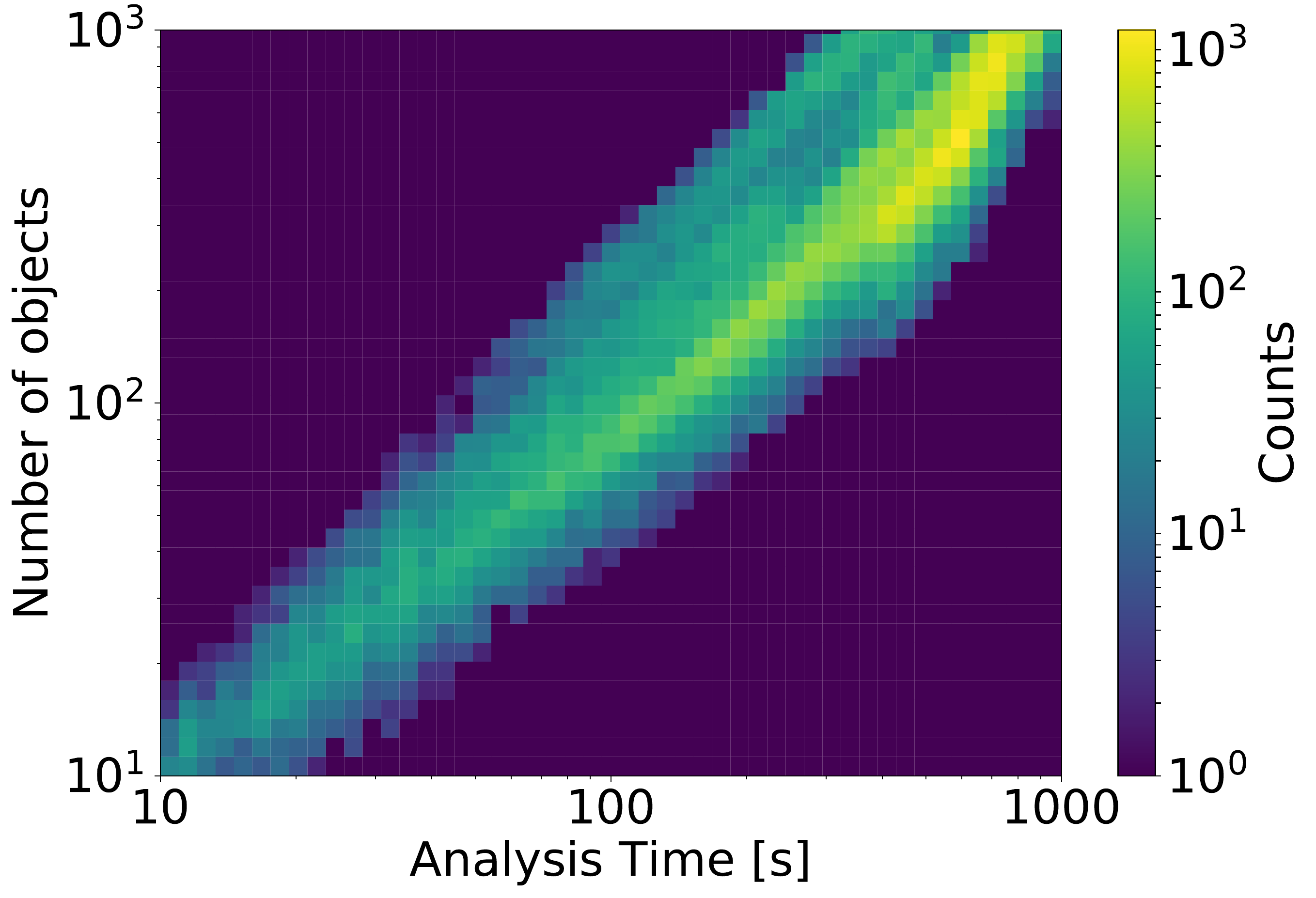}
\includegraphics[width=3.6in]{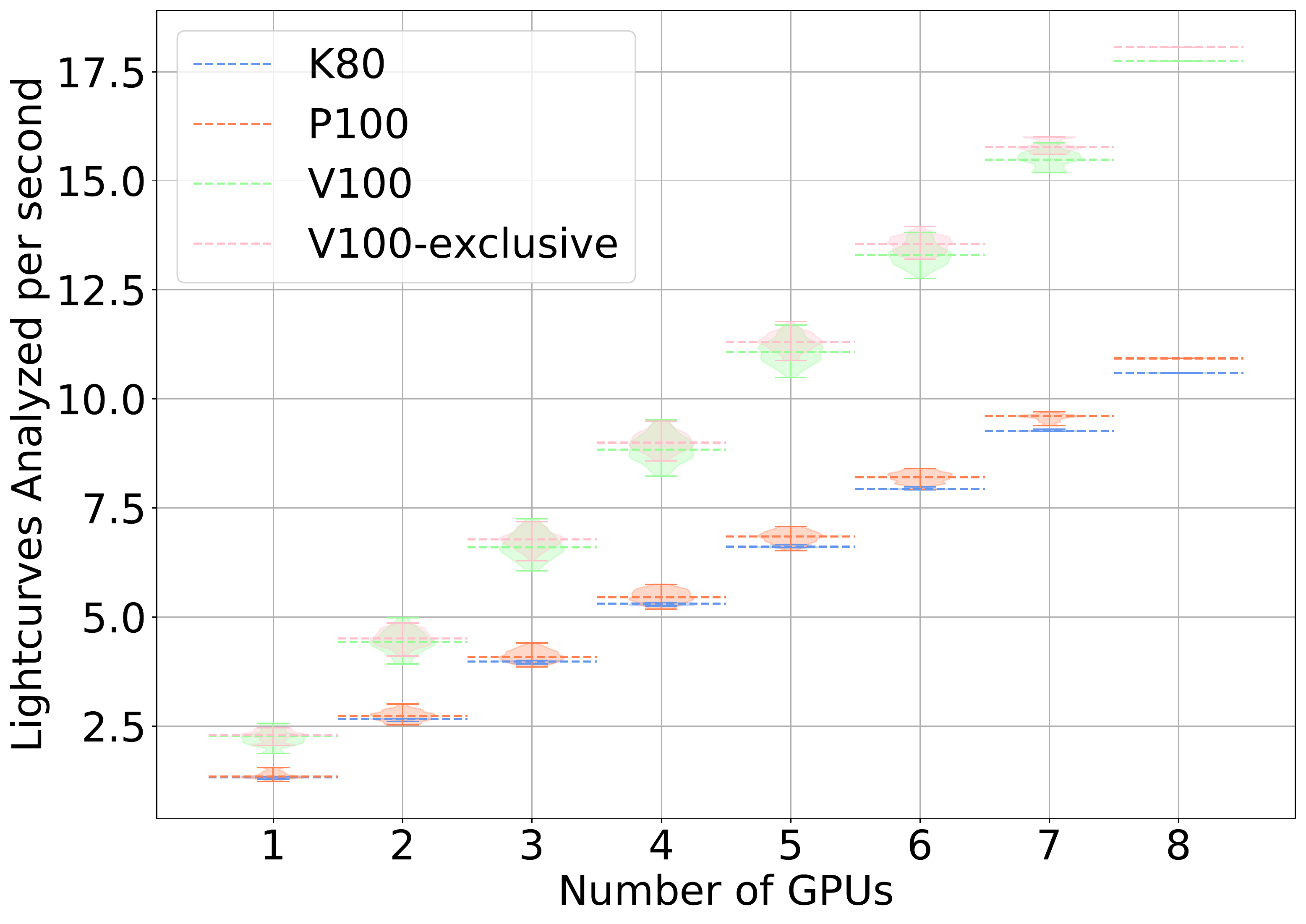}
\caption{Left: Two dimensional histogram of the number of objects analyzed vs. analysis time for the ZTF light curves. Right: Scaling data based on GPU resources for the algorithm discussed in the text. We note that the V100 and V100-exclusive analysis (where jobs are restricted to run on their own on the GPU) are closely overlapping.}
\label{fig:period_finding_scaling}
\end{figure*}

\begin{figure}[t]
\includegraphics[width=3.4in]{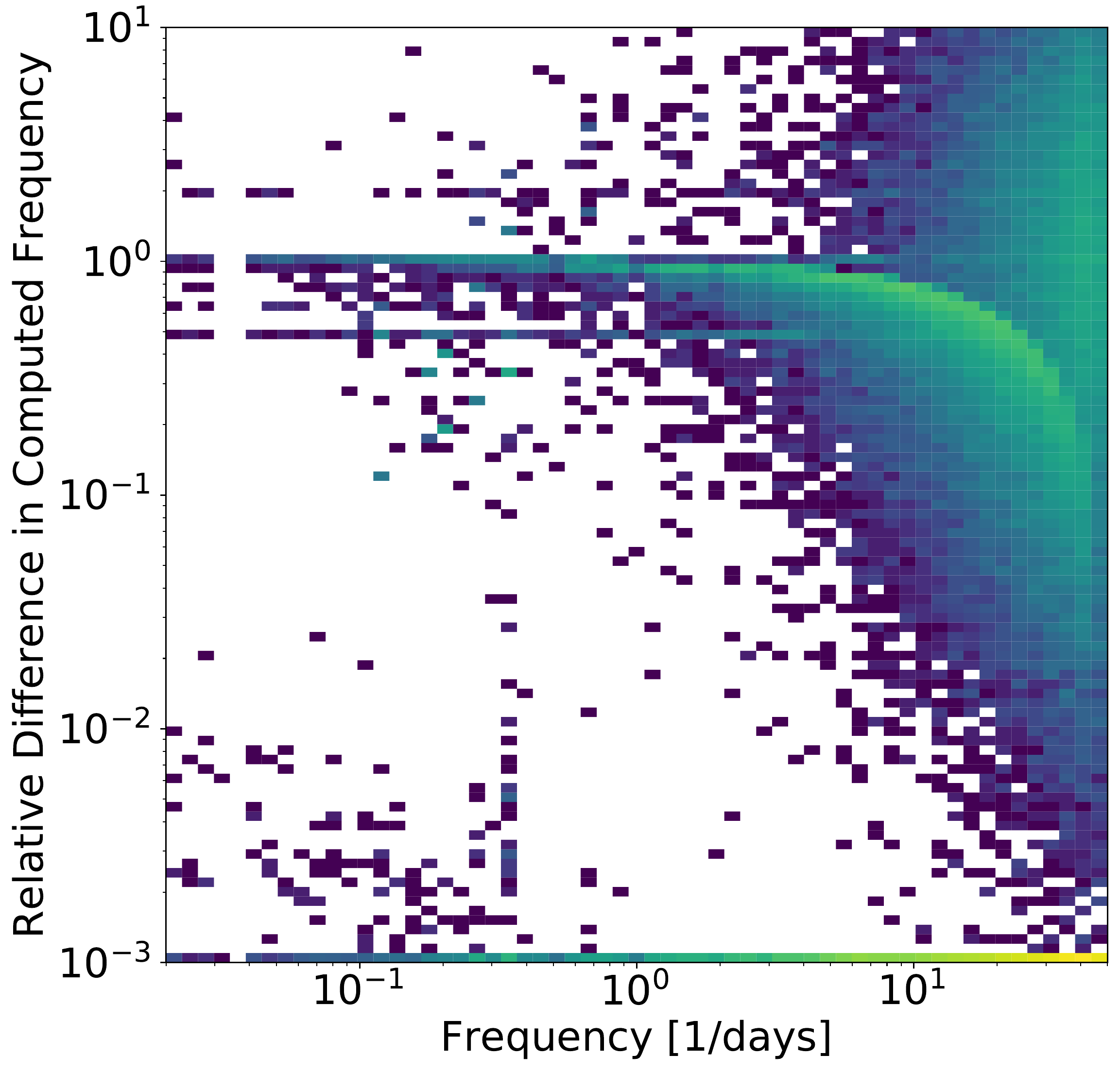}
\caption{Two dimensional histogram of the relative difference between an analysis of variance computation with oversampling factors of 3, as used in the main analysis, and an oversampling factor of 10 vs. highest significance period identified by the oversampling of 10 analysis.}
\label{fig:periods_relative}
\end{figure}

The software \texttt{gce} is a \texttt{Python} code that prepares light curves and their magnitude information for input into the Cython-wrapped CUDA program. 
Therefore, the user interface is entirely \texttt{Python}-based. 
The period range searched varies between 30\,min and half of the baseline, $T$; 30\,min was chosen as a lower bound for computational considerations. We use a frequency step $df$ of $\frac{1}{3\,\times T}$, oversampled by a factor of 3 in order to account for the irregular sampling. This oversampling term was measured empirically by trying the frequency grid on a variety of test cases, and shown to be the minimum factor that gave the correct period for a small sample of eclipsing binaries and RR Lyrae as determined by a higher resolution analysis. To evaluate the efficacy of this choice, Figure~\ref{fig:periods_relative} shows a two dimensional histogram of the relative difference between an analysis of variance computation with an oversampling factor of 3, as used in the main analysis, and an oversampling factor of 10. We plot this relative difference as a function of the highest significance period identified by the oversampling factor of 10 analysis. We plotted any relative difference values below that of $10^{-3}$ in the bottom row of the histogram, indicating agreement at the 0.1\% level. The main feature in this histogram, beyond the build-up of support of equal periods at the bottom, is the approach of a curve to a relative difference of 1.0, indicating those sources that disagree by a factor of 2 are modulated by the effect of diurnal sampling. There is a less dense horizontal line at those sources at half of the frequency. This analysis indicates that a more refined period grid, perhaps as a secondary step in the analysis, may be useful in the future. We note that phasing agreement to better than the 0.1\% level is already achieved, and accuracies at this level are required to, for example, detect small but detectable period changes present in a variety of astrophysical processes.

We note that this baseline will change for each chunk analyzed; the frequency array is then simply $f = f_{min} + df \times [0...\mathrm{ceil}((f_{max} - f_{min}) / df)]$. We use 20 phase bins and 10 magnitude bins in the conditional entropy calculation, which amounts to the size of the phase-folded, two dimensional histogram. We note that conditional entropy, as a 2D binning algorithm, does not currently have weights implemented. In principle, this could be done similarly to the case of Lomb-Scargle below, although careful consideration should be taken for how to handle, for example, eclipsing systems where downweighting fainter points with correspondingly larger error bars could be detrimental to identifying eclipses.

The second algorithm is a Lomb-Scargle (LS, \citealt{Lo1976}; \citealt{Sc1982}) implementation; these algorithms are similar to Fourier Transforms, but for irregularly sampled data. 
In this version, it decomposes the time series into the frequency domain using a linear combination of sine waves $y = a \cos \omega t + b \sin\omega t$. 
If we define $T$ as the period with the angular frequency $\omega = 2\pi T$, the periodogram is defined as:
\begin{multline*}
P(\omega) = \frac{1}{2\sigma^2}\left\{\frac{\left[\sum^N_{n=1} w_n(m_n - \bar{m})\cos\left[\omega(t_n-\tau)\right]\right]^2}{\sum^N_{n=1}\cos^2\left[\omega(t_n-\tau)\right]} \right.\\
\left. + \frac{\left[\sum^N_{n=1} w_n(m_n - \bar{m})\sin\left[\omega(t_n-\tau)\right]\right]^2}{\sum^N_{n=1}\sin^2\left[\omega(t_n-\tau)\right]}\right\},
\end{multline*}
where 
\begin{equation}
w_n = \frac{\frac{1}{\sigma_n^2}}{\sum^N_{n=1} \frac{1}{\sigma_i^2}}; \tau = \tan(2\omega \tau) = \frac{\sum^N_{n=1} \sin(2\omega t_n)}{\sum^N_{n=1} \cos(2\omega t_n)}.
\end{equation}
The algorithm we use is also implemented on GPUs \footnote{https://github.com/johnh2o2/cuvarbase}.

\begin{figure}[t]
\includegraphics[width=3.5in]{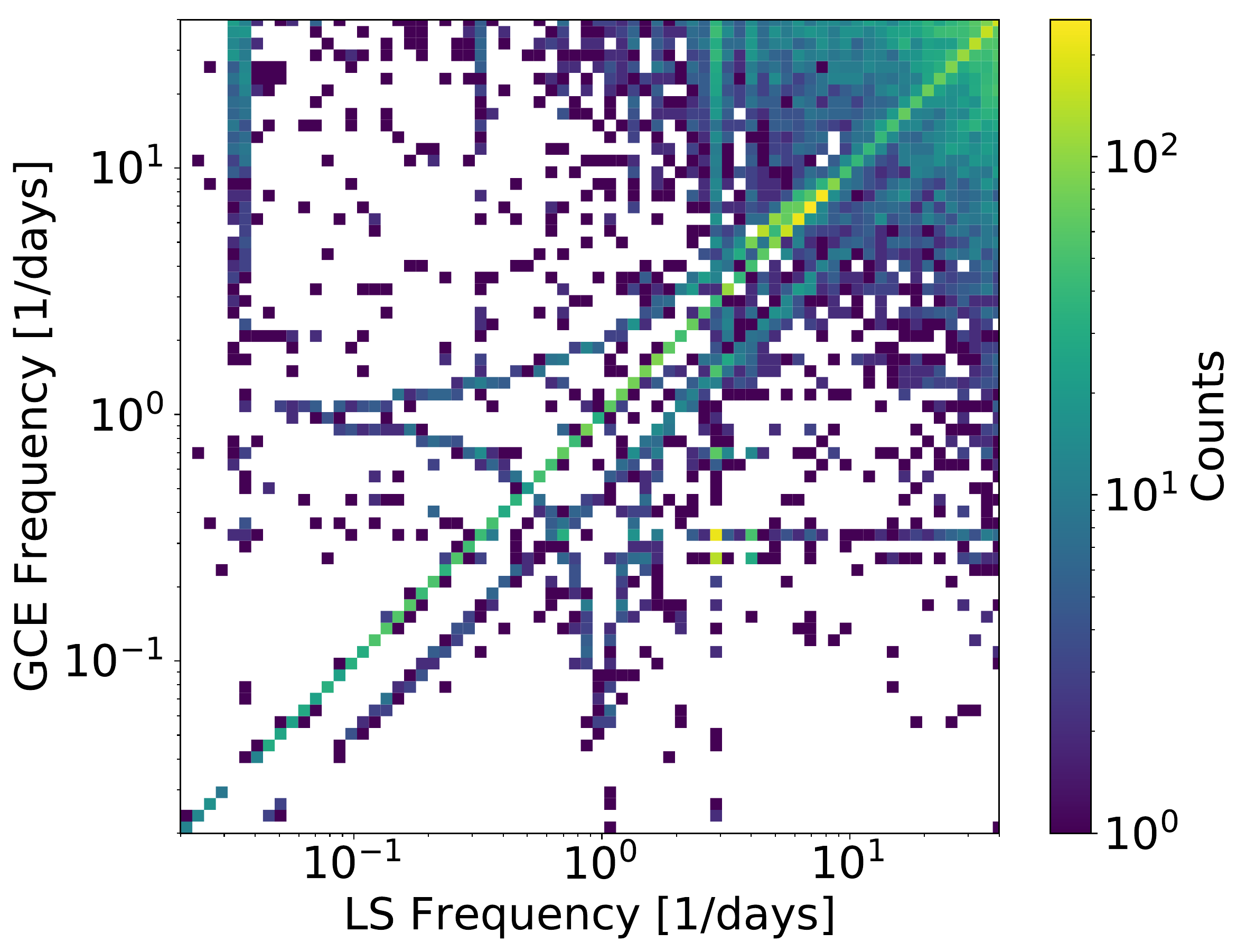}
\caption{Two dimensional histogram comparing LS (x-axis) and GCE (y-axis).}
\label{fig:periods_GCE_LS}
\end{figure}

\begin{figure*}[t]
\includegraphics[width=3.5in]{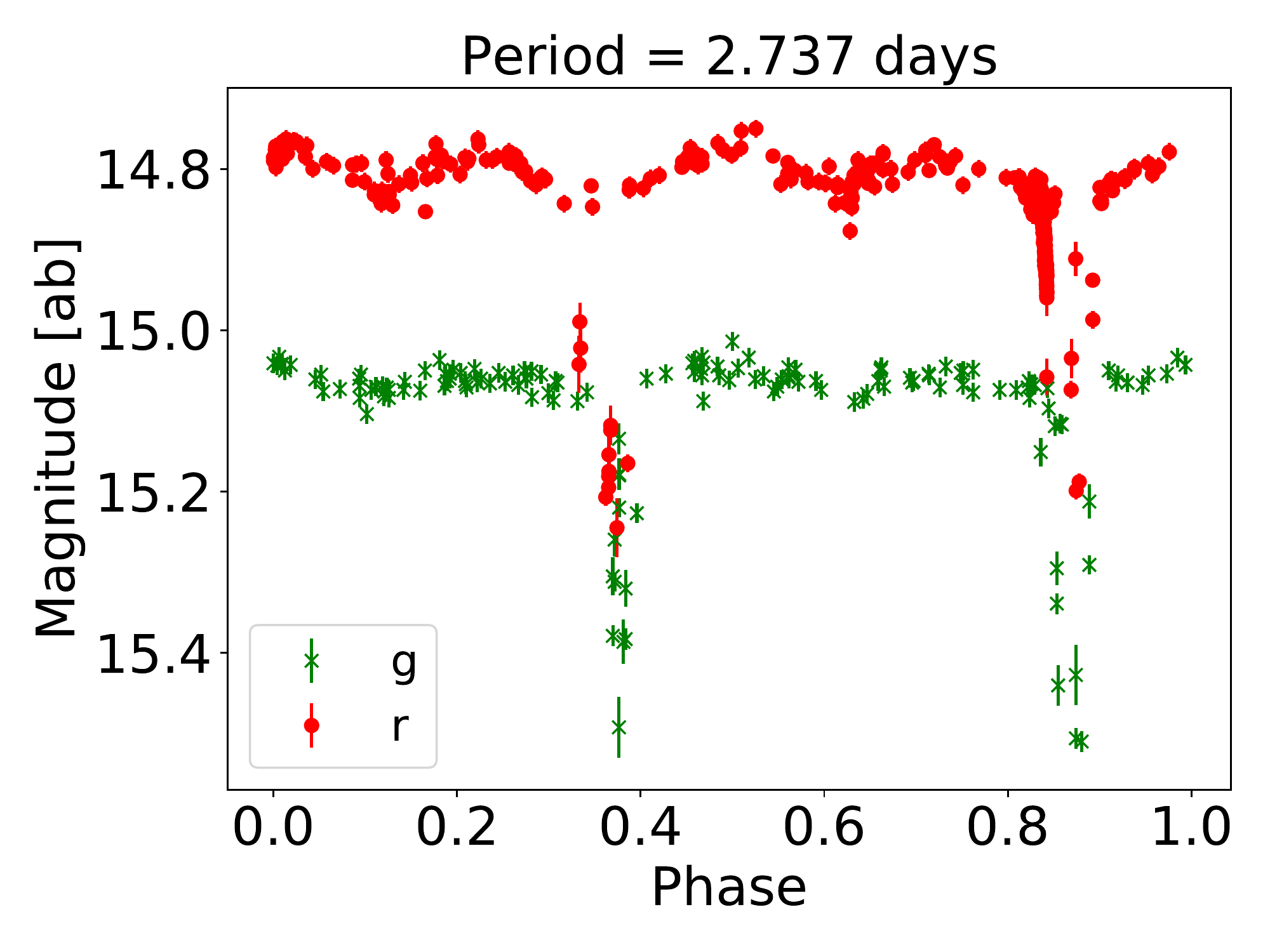}
\includegraphics[width=3.5in]{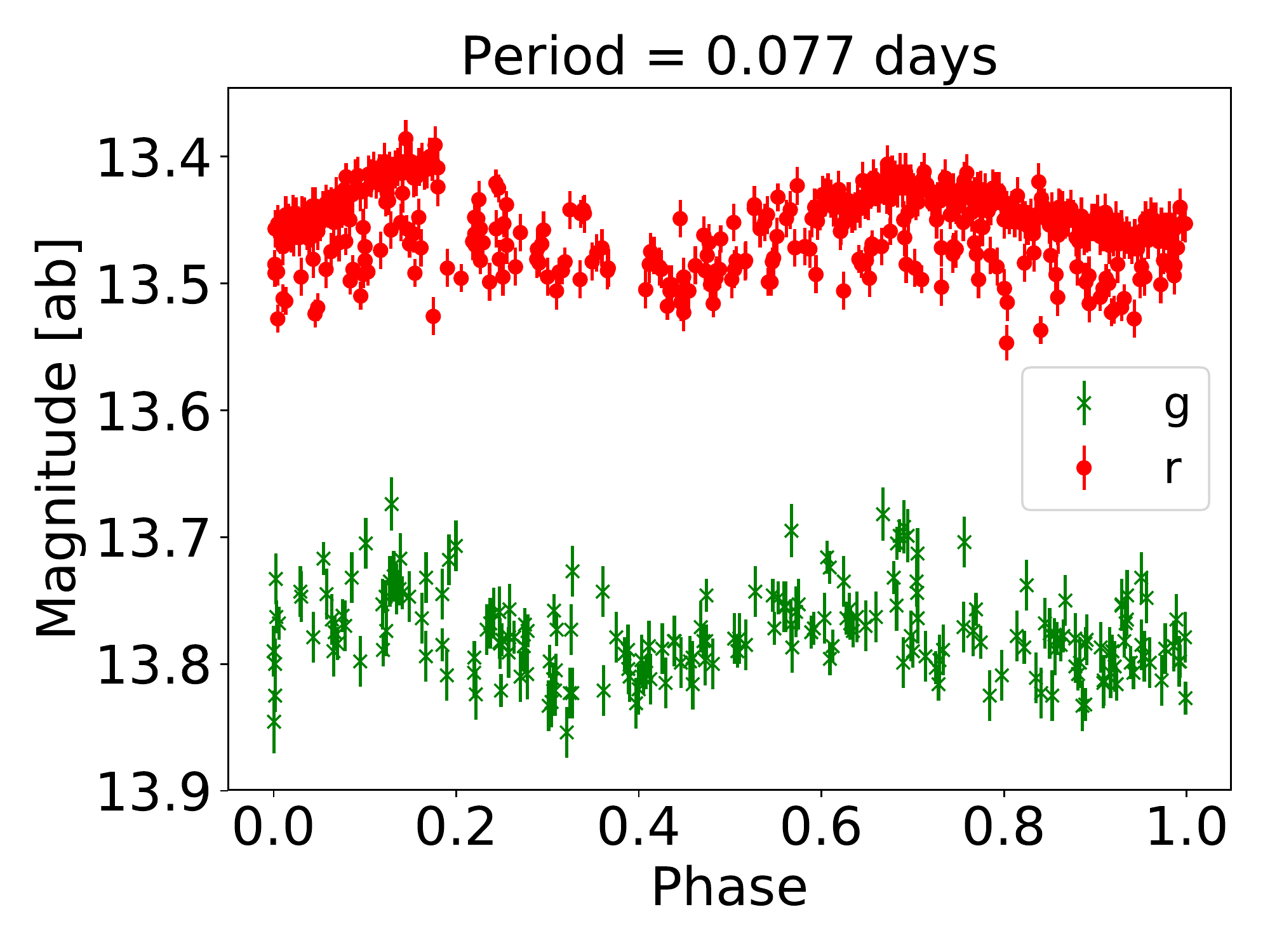}
\caption{Example periodic variables (their estimated periods are given in the titles), identified by GCE (left) and LS (right) as high significance, while the other period finding algorithm finds marginal significance.}
\label{fig:examples_GCE_LS}
\end{figure*}

Figure~\ref{fig:periods_GCE_LS} shows a two dimensional histogram comparing LS (x-axis) and GCE (y-axis).
For many objects, the peak frequencies identified are in agreement, with a subset identified as differing by half or twice the period.
When this occurs, LS preferentially finds the shorter period, which occurs in a number of common scenarios. This includes when analyzing eclipsing binaries whose primary and secondary eclipses do not differ greatly in depth; in this case, LS tends to find a period equal to half the true value, while CE will find the correct value.
For a further subset, the peak frequencies identified are different, requiring, in principle, a tie-breaker algorithm to determine the ``best'' choice.
A further, interesting feature in these histograms are curved ``lines,'' symmetrical about the diagonal, that have vertical and horizontal asymptotes at periods corresponding to one day, which arise from the true period ``beating'' with a one day period arising from diurnal aliasing.

\begin{figure}[t]
\includegraphics[width=3.5in]{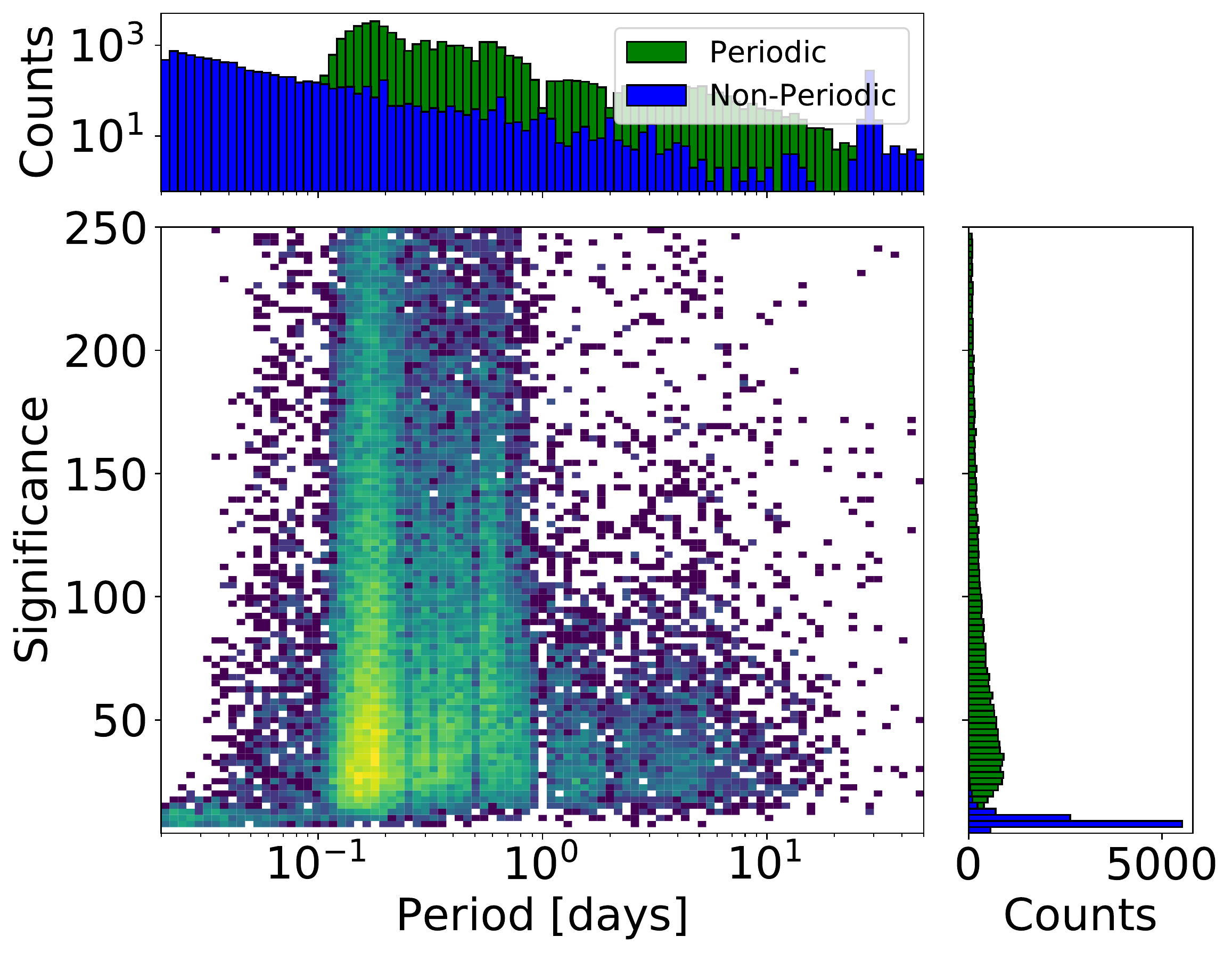}
\caption{Two dimensional histograms of the significance vs. period for the ``periodic'' set of objects (van Roestel et al. in prep). The one dimensional histograms are marginalized versions of the two dimensional histogram (green). For comparison, we include a ``non-periodic'' set of objects in blue (van Roestel et al. in prep).}
\label{fig:period_significance_periodic}
\end{figure}

Figure~\ref{fig:examples_GCE_LS} shows example periodic variables identified by GCE and LS as high significance, while the other period finding algorithm finds marginal significance; the GCE example has both eclipsing and underlying modulating behavior, more difficult for LS to recover, while the LS example identifies low-amplitude modulation in otherwise noisy data, difficult for GCE to recover within its 2-D histograms.
To address this issue, we take the top 50 frequencies identified by each of the algorithms, with no use of harmonic summing or otherwise period combining techniques, and run each through a CPU-based multi-harmonic analysis of variance (AOV, \citealt{ScCz1998}) code in the 200 frequency bins, covering a frequency range between peak frequency $-$ 100$\times$\,df to peak frequency $+$ 100$\times$\,df \footnote{https://github.com/joshuazd/lssfds}. To determine the best period, we evaluate their ad-hoc ``significance'' using the statistic array $s$, which is the same length as the frequency array:
\begin{equation}
    \mathrm{significance} = \left(\mathrm{max(s)}-\mathrm{mean(s)}\right)/\mathrm{std(s)}.
\end{equation}
We note that $s$, which is parameterized by the period array, corresponds to, for example, the entropy in the phased 2D histograms for conditional entropy, or an estimate of the Fourier power for LS.
While other examples of ``significance'' are also possible, such as approximate significance estimates for LS \citep{Bal2008}, we find this is sufficient for simply rank-ordering the frequencies across all algorithms and make simple comparisons between objects.
We also point out that this technique throws away information from sub-dominant periods in a light curve, and so for objects where multiple periods are important, this technique will be suboptimal.
We can evaluate the threshold for significance appropriate for evaluating a confident periodic source; in Figure~\ref{fig:period_significance_periodic}, we plot the significance vs. period for the ``periodic'' set of objects (van Roestel et al. in prep). For comparison, we show marginalized one-dimensional histograms for both the periodic and non-periodic sets of objects, which show distinct differences in both. The non-periodic set show a distinct peak at low frequencies and around the lunar cycle, while the periodic set peaks distinctly in the range 0.1--1\,days, due to the high rate of RR Lyrae and Delta Scuti variables in this set. For scanning purposes, we can also compare the distribution of significances for these sets; for example, the 2nd percentile for periodic objects is $\sim$\,14.4, which corresponds to the 95th percentile of non-periodic objects. This would mean that a significance threshold of 14.4 would yield a false dismissal probability of 2\% while having a false alarm probability of 5\%. For further comparison, the relative rate of periodic to non-periodic sources is $\sim$\,0.13\%.

There were two main advantages to this otherwise complicated method.
The first is that there are known advantages to both CE and LS \citep{GrDr2013}, with one particularly successful for generic light curve shapes and the second for low-amplitude variables, as demonstrated in Figure~\ref{fig:examples_GCE_LS}.
To provide a single set of metrics based on these two algorithms, we needed a way to ``choose'' between CE and LS, from which came the choice of AOV, which has some of the benefits of both (generic light curve shapes with the potential for multiple harmonics). 
These motivates the second point, which is that the CPU-based version of AOV across the whole frequency grid would be computationally intractable, and so some of the benefits of the most suitable period finding algorithm are achieved despite its computational expense.
The GPU implementations of CE and LS were essential here, as CPU implementations of CE, AOV, and LS take $\sim$\,10\,s per light curve each, two orders of magnitude slower than their GPU counterparts.
We highlight a few example periodic objects coming out of this analysis in Figure~\ref{fig:examples}.

We show the analysis speed as a function of number of objects on the left of Figure~\ref{fig:period_finding_scaling}. 
As can be seen, the analysis takes $\sim$\,1 second per light curve across the range of light curves analyzed at a single time ($\sim$\,100 to $\sim$\,1000 samples).
At this point, we are dominated by the CPU cost of the AOV analysis, having put all other expensive computations on the GPU. Design of a GPU-based AOV analysis is ongoing, as well as breaking out the AOV process into separate processes (perhaps for objects that meet specific variability requirements).
We now perform scaling tests for the numerical setup described above. The right of Figure~\ref{fig:period_finding_scaling} shows the scaling test results on 10 GPUs running concurrently. 
These tests have been performed at a number of facilities: (i) the Minnesota Supercomputing Institute, with both K40 (11\,GB of RAM) and V100 GPUs (16\,GB of RAM), (ii) XSEDE's SDSC COMET cluster \citep{ToCo2014} with K80 (12\,GB of RAM) and P100 GPUs (16\,GB of RAM), and (iii) NERSC's CORI cluster with V100's (16\,GB of RAM).
As expected, the number of analyzed light curves is linear with the number of GPUs allocated. 

\begin{figure*}[t]
\includegraphics[width=3.5in]{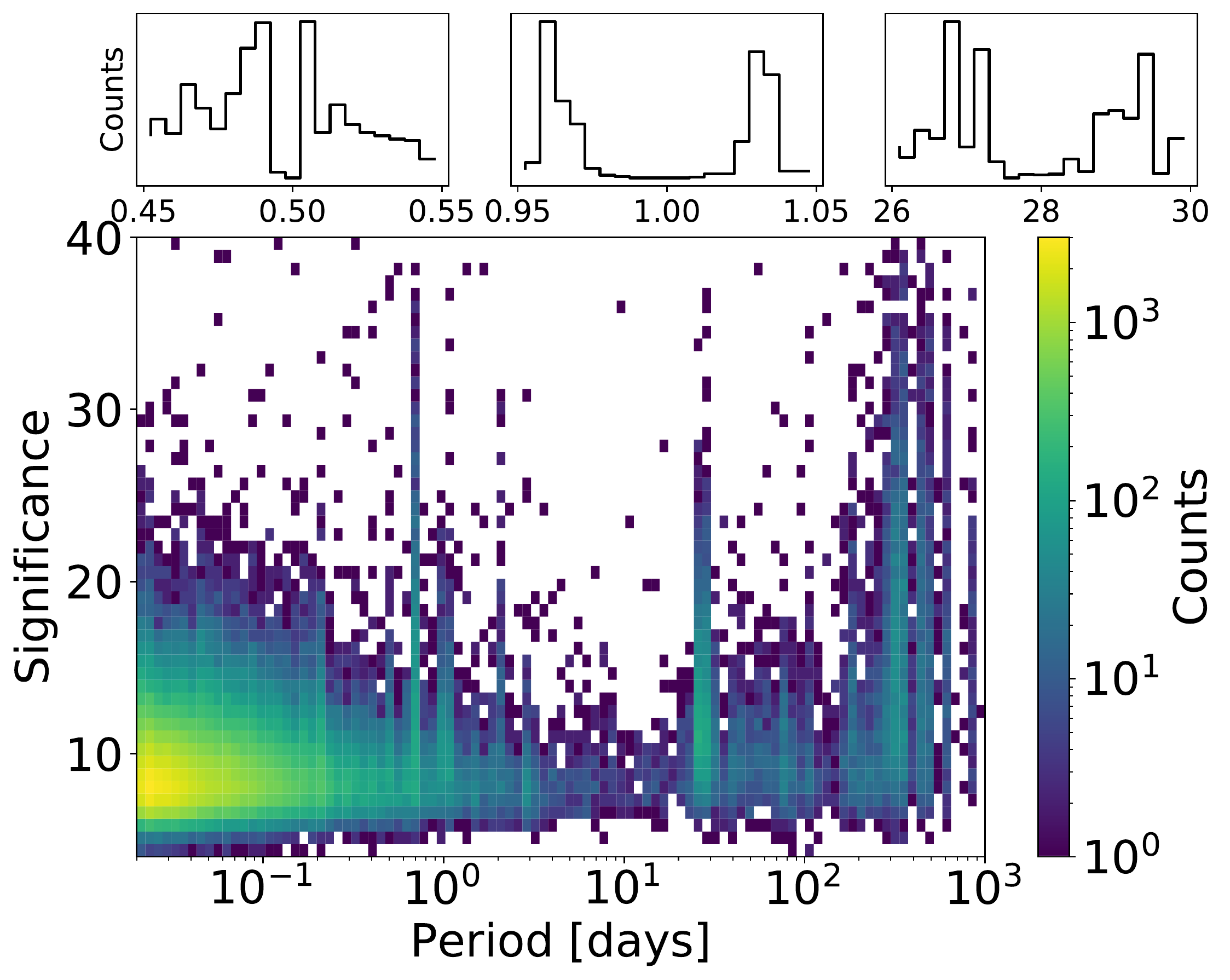}
\includegraphics[width=3.5in]{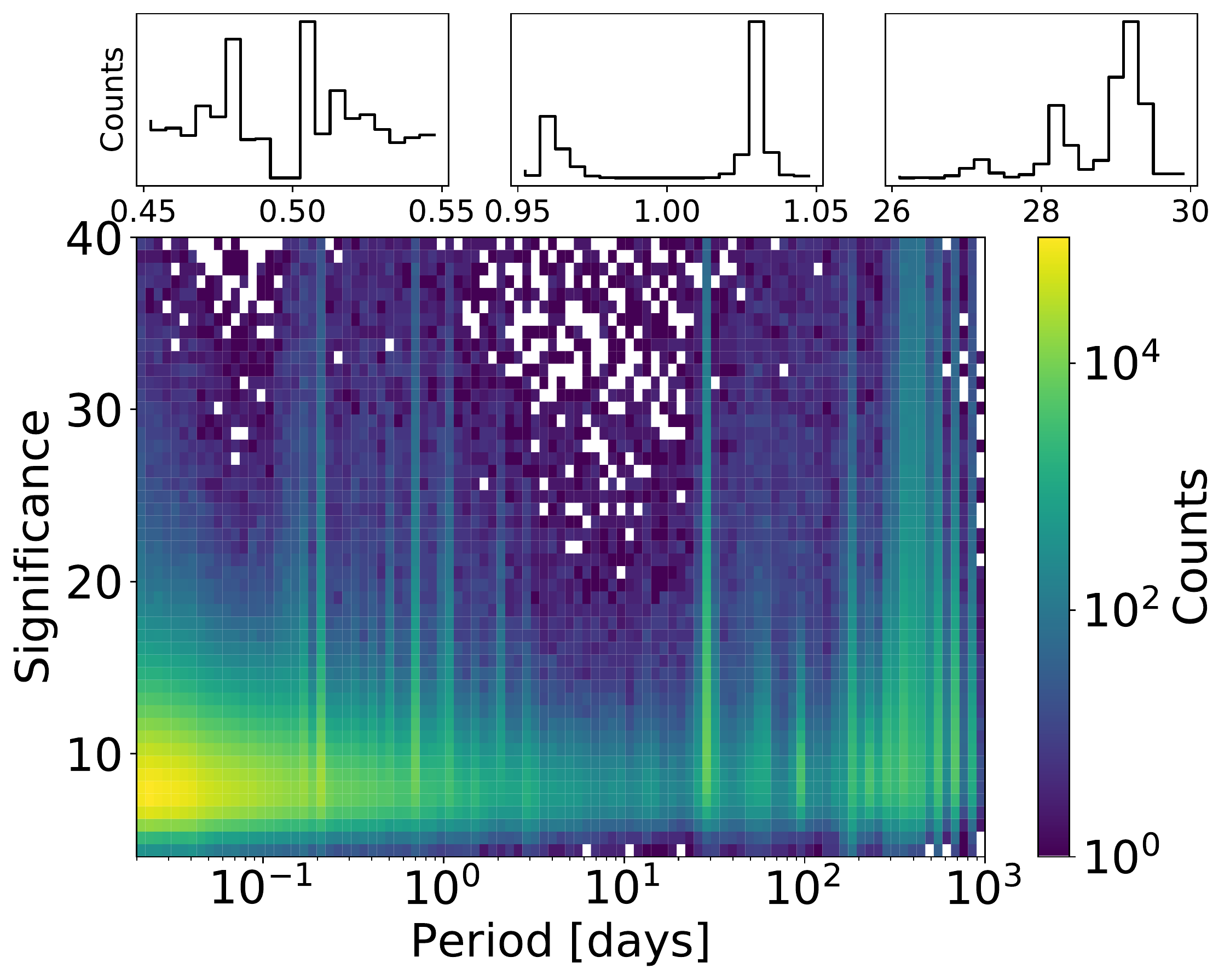}
\includegraphics[width=3.5in]{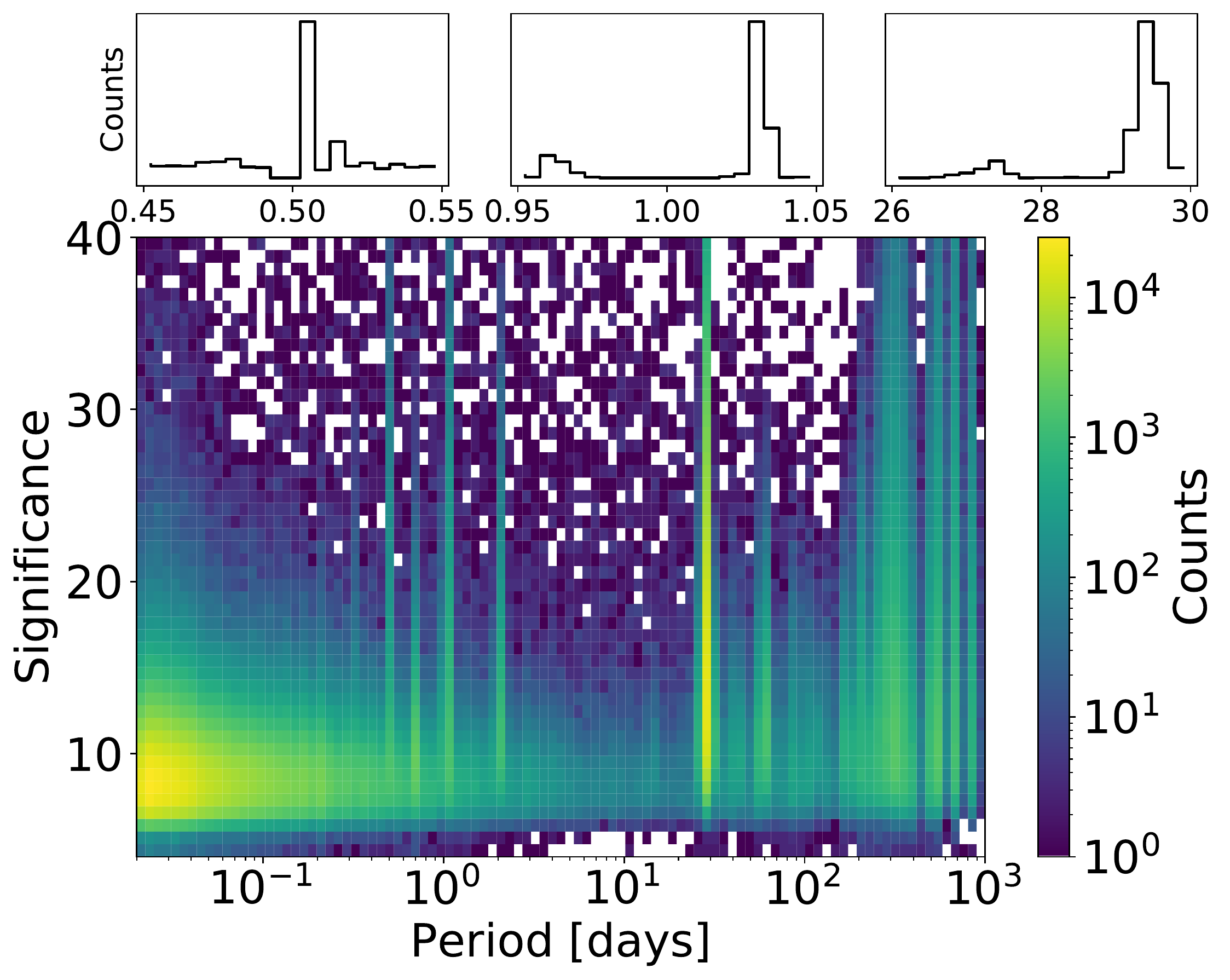}
\includegraphics[width=3.5in]{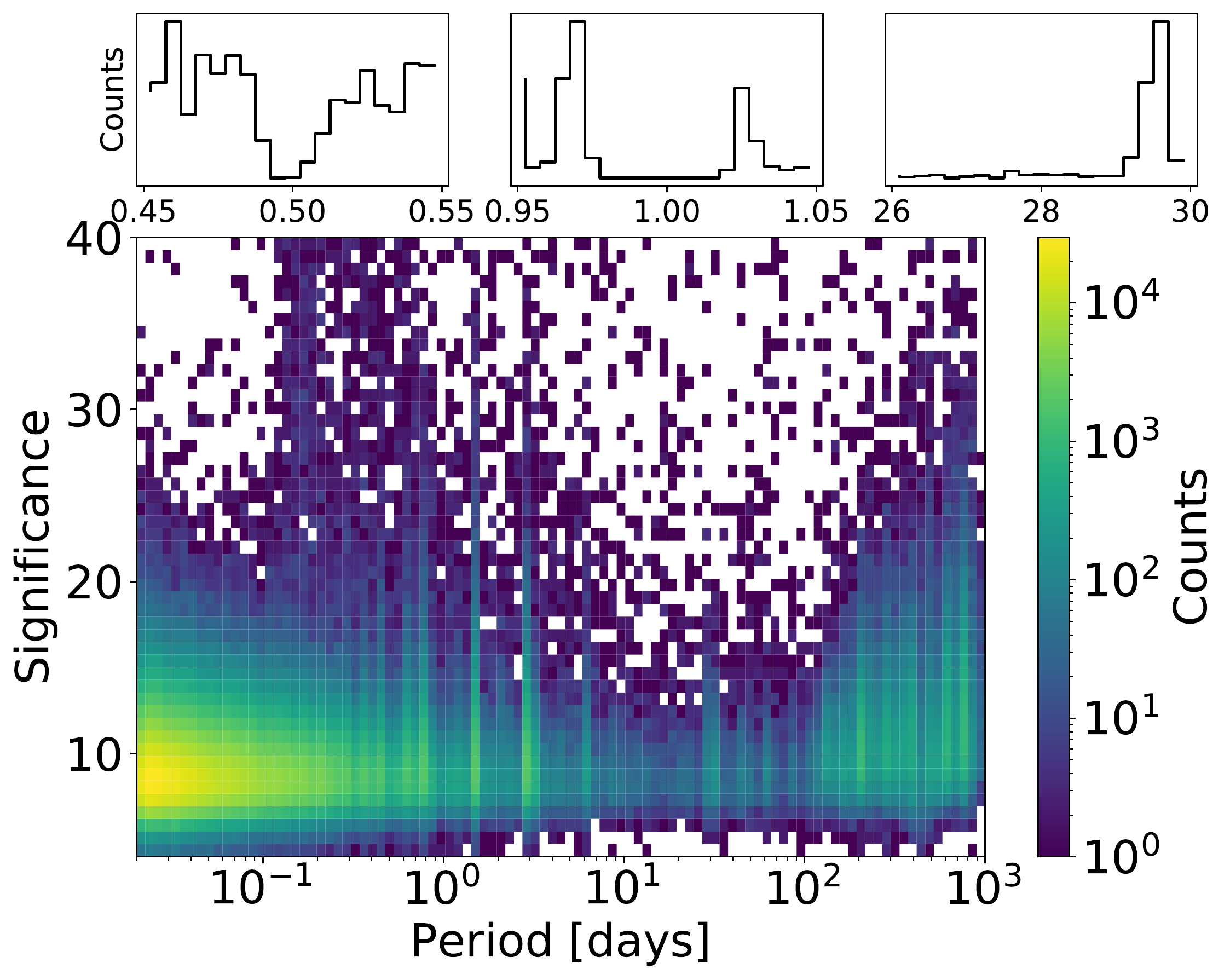}
\caption{Two dimensional histograms of the period significance based on the analysis of variance computation vs. highest significance period identified for field 296 (top left), 487 (top right), 682 (bottom left) and field 853 (bottom right). The one dimensional histograms focus on periods of 0.5\,days and 1\,days,to evaluate the effect of diurnal sampling on aliasing, and 28\,days, to evaluate the same for the lunar cycle.}
\label{fig:period_significance}
\end{figure*}

Figure~\ref{fig:period_significance} shows the two dimensional histogram of the period significance based on the analysis of variance computation vs. highest significance period identified for field 296 (top left), 487 (top right), 682 (bottom left) and field 853 (bottom right). We can identify a number of cases where a non-astrophysical signal is found due to the sampling pattern. It is clear based on the histograms that fractions and multiples of a sidereal day, as well as near sidereal months with the lunar cycle, induce false period estimates. We remove at the period finding level bands around fractions and multiples of a sidereal day to help mitigate this. Specifically, we remove frequencies (in cycles/day) of 47.99--48.01, 46.99--47.01, 45.99--46.01, 3.95--4.05, 2.95--3.05, 1.95--2.05, 0.95--1.05, 0.48--0.52 and 0.03--0.04; this removes $\sim$\,1\% of the frequency range. We chose these frequencies based on an iterative procedure; we would evaluate a subset of objects, examine histograms of period excesses identified at these frequencies, remove them, and then reevaluate. 
This will be convolved with longer term trends, including seasonal and annual variations. We note that removal of these frequencies may have removed the true variables covering these period ranges. The marginalized histograms in Figure~\ref{fig:period_significance} indicate that a small excess of sources at the edge of the removed frequency bands remains due to the diurnal and lunar aliases, showing that a broader removal could be useful.

\begin{figure*}[t]
\includegraphics[width=3.5in]{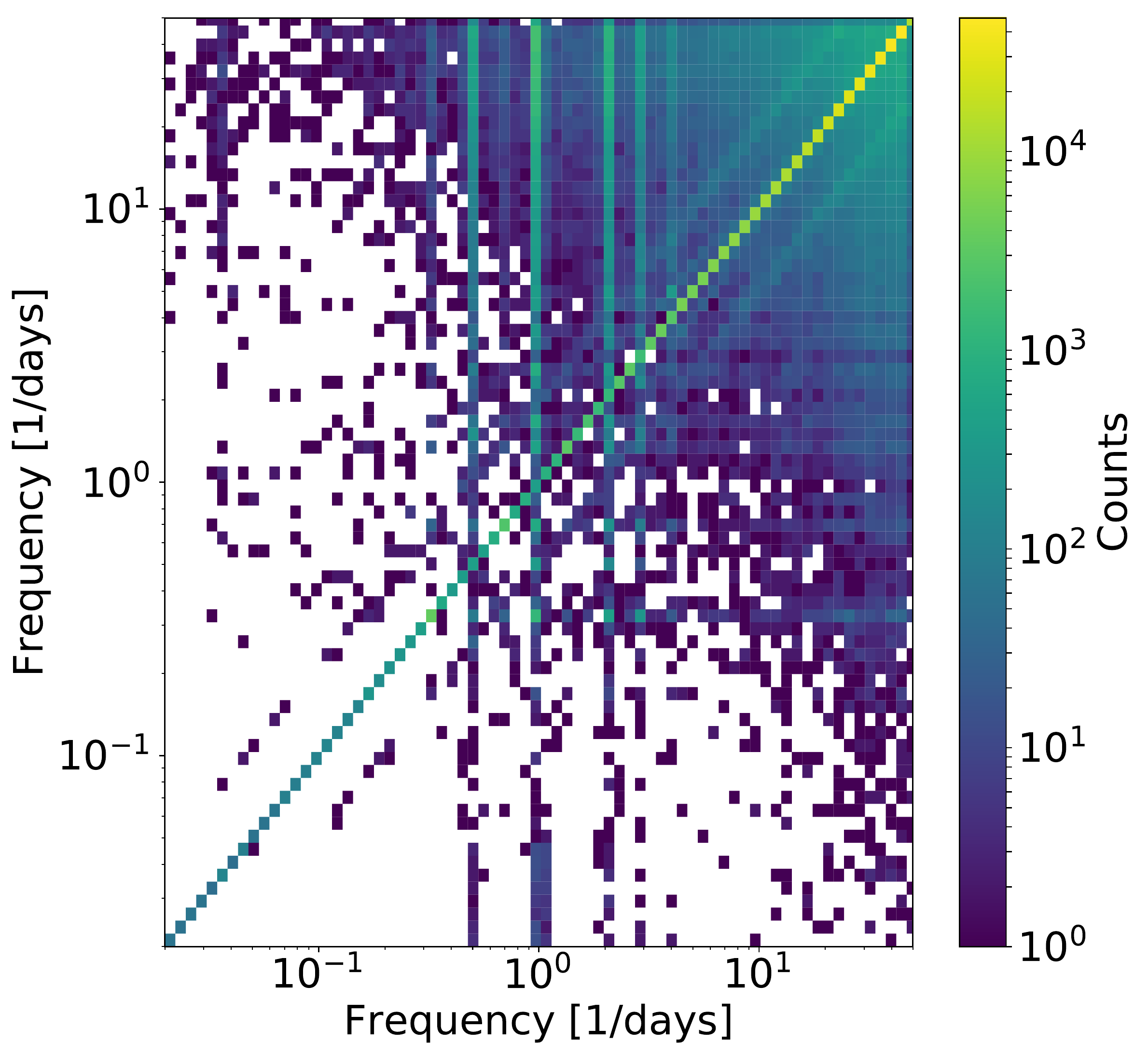}
\includegraphics[width=3.5in]{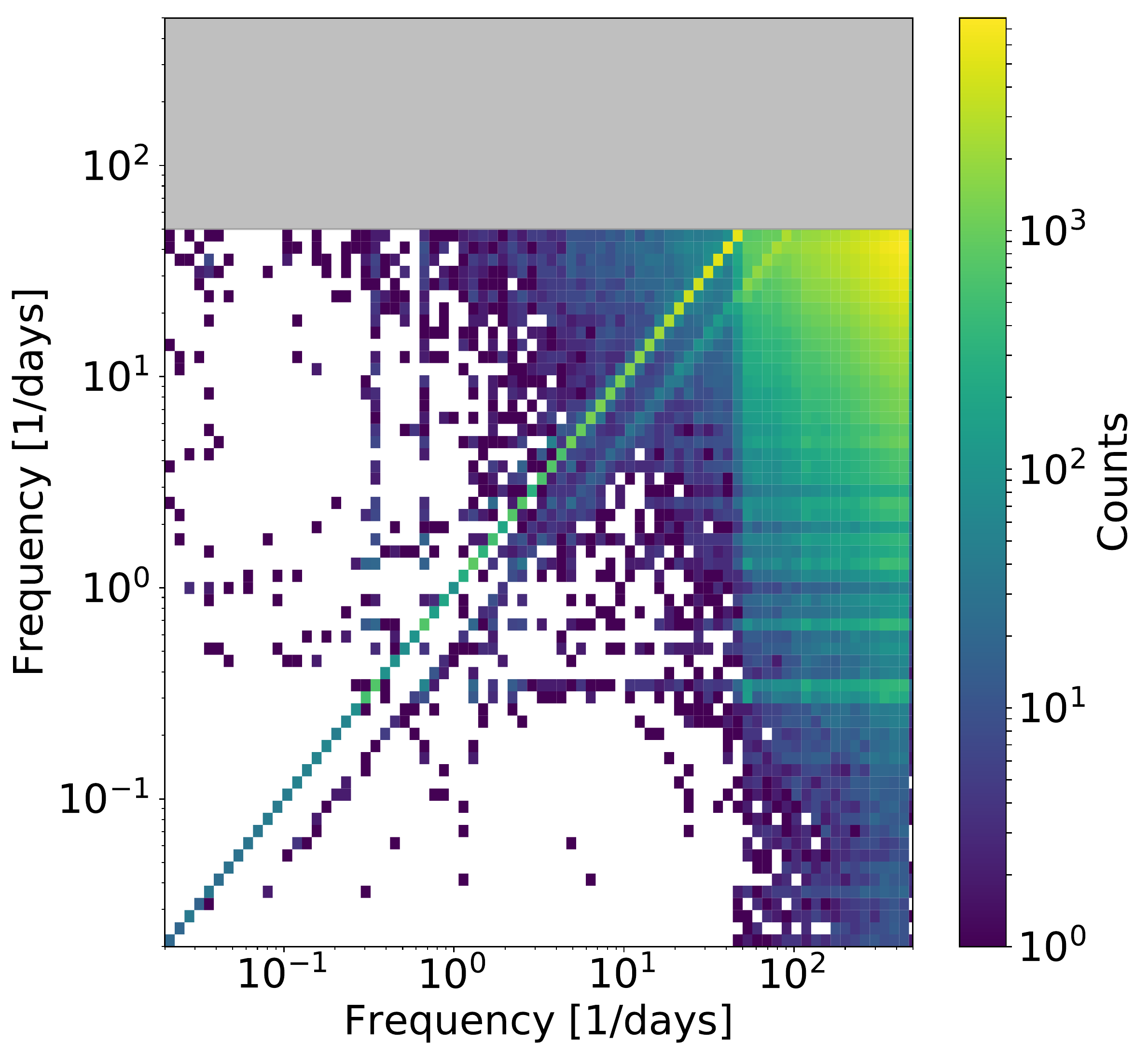}
\includegraphics[width=3.5in]{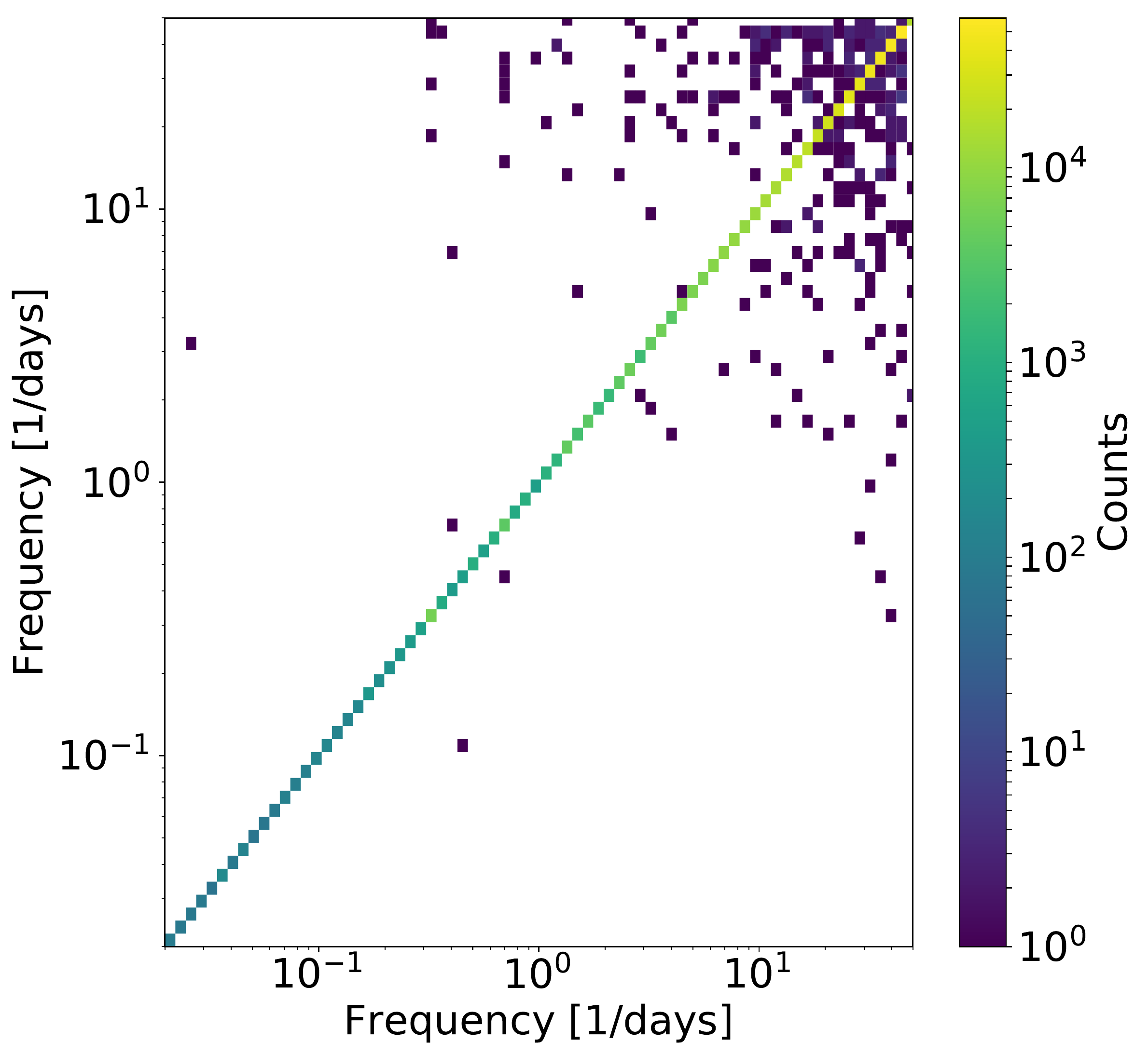}
\includegraphics[width=3.5in]{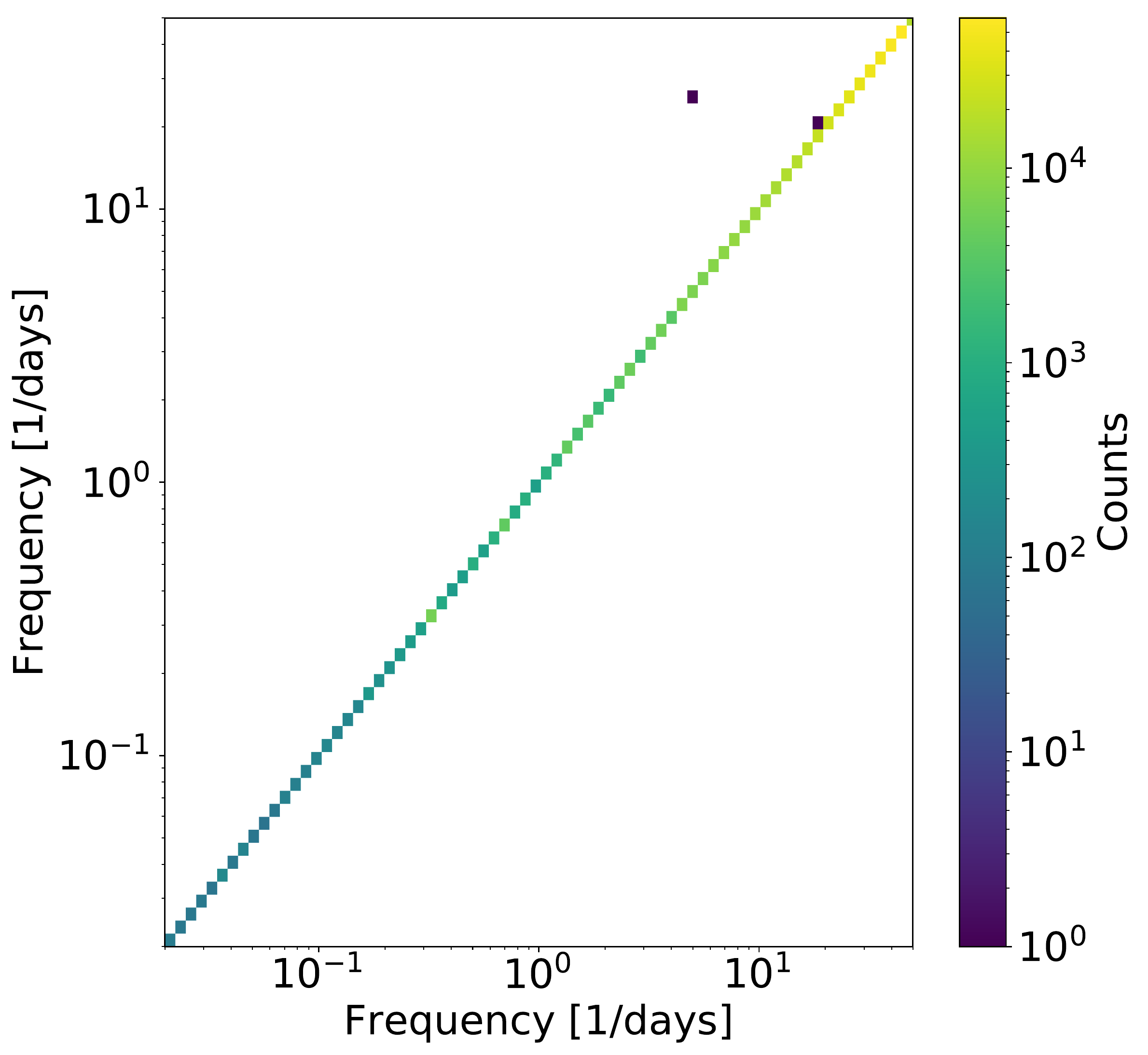}
\caption{On the top left is a comparison between two period finding analyses of field 700 comparing one without the aliasing-dominated frequency bands removed (x-axis) and the typical setup (y-axis). On the top right is the same where the x-axis version has a period range searched down to 3\,min. The gray region corresponds to a period region not available to that run. On the bottom row, we compare runs where we use the typical setup (left) and one where the GPU-based period finding is run 3 times (right).}
\label{fig:periods_comparison}
\end{figure*}

The top left of Figure~\ref{fig:periods_comparison} shows a comparison between period finding analyses of field 700 using the setup used in this analysis and one where we do not remove these bands; in general, most of the objects are assigned the same period. In addition to some differences due to GPU errors (see below) and the differences that arise in the significance calculation where different frequencies reach the AOV stage, a significant fraction receive periods in narrow frequency bands that are otherwise cut, indicating the importance of these cuts.
The top right of Figure~\ref{fig:periods_comparison} shows a comparison between period finding analyses of field 700 using the setup used in this analysis and one where periods are searched down to 3\,min; similar to before, in general, many of the objects are assigned the same period. Some objects are assigned a period that corresponds to the second harmonic of the original analysis. Also, some objects have very short periods assigned, as typically occurs for marginal periodic signals.

We also characterized GPU-based transient errors that affect a small fraction of objects when running on the HPC clusters used here \citep{TiGu2015}.
The bottom row of Figure~\ref{fig:periods_comparison} shows a comparison between two identical period finding analyses of field 700 using the setup used in this analysis and one where the GPU-based period finding is run 3 times; while two outliers remain, this is enough to clean up the distributions at the cost of computational efficiency.
We expect that if there are algorithmic speed-ups enabling this to be performed feasibly over the entire data set, it will be useful to mitigate this issue.

\section{Conclusion}

In this paper, we introduce the variability metrics and the periodicity algorithms that are used to derive ZTF's variable star catalog based on ZTF's DR2. 
In addition to the variability metrics suggested by \cite{PaSo2017}, we included robust measurements of the light curve amplitude and a Fourier-decomposition to characterize the shape of the folded light curves.
We designed a hierarchical period-finding scheme that utilized multiple period-finding algorithms in order to be as sensitive as possible across the period parameter space of amplitude and period.
This analysis provides the input statistical features that will provide training sets for the machine learning based catalogs, presented in a partner publication (van Roestel et al. in prep).
This publication will present classifiers and associated metrics that evaluate the variability, periodicity, completeness, purity and classifications for these objects.
To create the catalog, \texttt{ztfperiodic} has been applied to a variety of large-scale computing systems and NVIDIA-based GPUs. Setting up and executing the proposed analysis required access to GPU arrays of this size, and the computational needs will only grow in future releases from ZTF and larger data sets from future surveys.

Going forward, the priority will be to analyze the data from the ZTF's Third Data Release (June 24, 2020)\footnote{\url{https://www.ztf.caltech.edu/page/dr3}}. 
Many of the choices made in this paper should be revisited for this process.
These include:
\begin{itemize}
\item Restricting to one point within 30 minutes (to remove high cadence data)
\item Restricting to periods greater than 30 minutes (to make computing tractable)
\item Restricting to a minimum of 50 points per light curve (for metric efficacy)
\item Period range exclusions (to remove effects from cadence aliasing)
\item Period finding choices, including the choice of the oversampling factor
\end{itemize}
Outside of more epochs yielding more light curves passing the 50 point cut, we expect to revisit the choices made in the period finding. Ongoing work includes devising faster algorithms to improve the scaling of the processing. Other options include combining photometric points from other surveys.

We also are exploring the benefits of ``clipping'' light curves for outlier removal, either using a comparison with a given $\sigma$ difference or percentile based cuts.
In general, we continue to improve the algorithms looking ahead, given that computational burdens are growing significant.
For LSST in particular, the number of variable sources expected is significantly higher, because of both its increased depth and $\sim$\,5 mmag precision, given that the fraction of variables increases as a power law with increasing photometric precision \citep{HuEv2006}. 
Translation of the work described here to LSST is obviously of great interest, as LSST will overlap with the faint end of ZTF, and have a limiting magnitude far beyond what ZTF can reach, meaning that it can access a far larger volume of sources. However, translating the work described here to LSST will be accompanied by several challenges. Obviously the cadence will be substantially lower due to LSST's smaller field of view relative to ZTF, which will impact primarily the ability to recover periods at the short end. Perhaps the biggest challenge is that LSST will not acquire sufficient samples to detect periodic behavior in many sources until several years into the survey, which will increase the baseline of the sampling relative to that of ZTF; this in turn means that a larger frequency grid must be searched to recover equivalent objects. Additionally, because LSST will contain far more sources than ZTF due to its depth, the computational cost will be significantly higher, as it is proportional to the total number of sources. 
This further motivates continued speed-ups of the algorithms presented here.

\section*{Data Availability}
The data underlying this article are derived from public code found here:\\ \href{https://github.com/mcoughlin/ztfperiodic}{https://github.com/mcoughlin/ztfperiodic}. DR2, including how to access the data, can be found at \href{https://www.ztf.caltech.edu/page/dr2}{https://www.ztf.caltech.edu/page/dr2}.

\acknowledgments
The authors would like to thank our referee, Dr. Aren Heinze, for very useful reports, improving the content of the paper.
M.~W.~Coughlin acknowledges support from the National Science Foundation with grant number PHY-2010970. M.~L.~K. acknowledges support from the National Science Foundation under grant DGE-0948017 and the Chateaubriand Fellowship from the Office for Science \& Technology of the Embassy of France in the United States.
MR has received funding from the European Research Council (ERC) under the European Union's Horizon 2020 research and innovation programme (grant agreement n759194 - USNAC).
DK is supported by NSF grant AST-1816492.

The authors acknowledge the Minnesota Supercomputing Institute\footnote{\url{http://www.msi.umn.edu}} (MSI) at the University of Minnesota for providing resources that contributed to the research results reported within this paper under project ``Identification of Variable Objects in the Zwicky Transient Facility.''
This research used resources of the National Energy Research Scientific Computing Center (NERSC), a U.S. Department of Energy Office of Science User Facility operated under Contract No. DE-AC02-05CH11231 under project ``Towards a complete catalog of variable sources to support efficient searches for compact binary mergers and their products.'' This work used the Extreme Science and Engineering Discovery Environment (XSEDE), which is supported by National Science Foundation grant number ACI-1548562. This work used the Extreme Science and Engineering Discovery Environment (XSEDE) COMET at SDSU through allocation AST200016.
M.L.K. also acknowledges the computational resources and staff contributions provided for the Quest/Grail high performance computing facility at Northwestern University.

Based on observations obtained with the Samuel Oschin Telescope 48-inch and the 60-inch Telescope at the Palomar Observatory as part of the Zwicky Transient Facility project. ZTF is supported by the National Science Foundation under Grant No. AST-1440341 and a collaboration including Caltech, IPAC, the Weizmann Institute for Science, the Oskar Klein Center at Stockholm University, the University of Maryland, the University of Washington (UW), Deutsches Elektronen-Synchrotron and Humboldt University, Los Alamos National Laboratories, the TANGO Consortium of Taiwan, the University of Wisconsin at Milwaukee, and Lawrence Berkeley National Laboratories. Operations are conducted by Caltech Optical Observatories, IPAC, and UW. 

\clearpage
\bibliographystyle{aasjournal}
\bibliography{references}

\begin{thebibliography}{}
\expandafter\ifx\csname natexlab\endcsname\relax\def\natexlab#1{#1}\fi
\providecommand{\url}[1]{\href{#1}{#1}}
\providecommand{\dodoi}[1]{doi:~\href{http://doi.org/#1}{\nolinkurl{#1}}}
\providecommand{\doeprint}[1]{\href{http://ascl.net/#1}{\nolinkurl{http://ascl.net/#1}}}
\providecommand{\doarXiv}[1]{\href{https://arxiv.org/abs/#1}{\nolinkurl{https://arxiv.org/abs/#1}}}

\bibitem[{Alcock {et~al.}(2000)Alcock, Allsman, Alves, Axelrod, Becker,
  Bennett, Cook, Dalal, Drake, Freeman, \& et~al.}]{AlAl2000}
Alcock, C., Allsman, R.~A., Alves, D.~R., {et~al.} 2000, The Astrophysical
  Journal, 542, 281–307, \dodoi{10.1086/309512}

\bibitem[{{Amaro-Seoane et al.}(2017)}]{AmAu2017}
{Amaro-Seoane et al.} 2017, {{Laser Interferometer Space Antenna}}.
\newblock \doarXiv{1702.00786}

\bibitem[{Antonini {et~al.}(2017)Antonini, Toonen, \& Hamers}]{AnTo2017}
Antonini, F., Toonen, S., \& Hamers, A.~S. 2017, The Astrophysical Journal,
  841, 77, \dodoi{10.3847/1538-4357/aa6f5e}

\bibitem[{Baluev(2008)}]{Bal2008}
Baluev, R.~V. 2008, Monthly Notices of the Royal Astronomical Society, 385,
  1279, \dodoi{10.1111/j.1365-2966.2008.12689.x}

\bibitem[{{Banerjee}(2018)}]{Ban2018}
{Banerjee}, S. 2018, \mnras, 473, 909, \dodoi{10.1093/mnras/stx2347}

\bibitem[{Behnel {et~al.}(2011)Behnel, Bradshaw, Citro, Dalcin, Seljebotn, \&
  Smith}]{Cython}
Behnel, S., Bradshaw, R., Citro, C., {et~al.} 2011, Computing in Science \&
  Engineering, 13, 31

\bibitem[{Bellm {et~al.}(2018)Bellm, Kulkarni, Graham, Dekany, Smith, Riddle,
  Masci, Helou, Prince, {et~al.}}]{Bellm2018}
Bellm, E.~C., Kulkarni, S.~R., Graham, M.~J., {et~al.} 2018, Publications of
  the Astronomical Society of the Pacific, 131, 018002,
  \dodoi{10.1088/1538-3873/aaecbe}

\bibitem[{Bellm {et~al.}(2019)Bellm, Kulkarni, Barlow, Feindt, Graham, Goobar,
  Kupfer, Ngeow, Nugent, Ofek, \& et~al.}]{BeKu2019}
Bellm, E.~C., Kulkarni, S.~R., Barlow, T., {et~al.} 2019, Publications of the
  Astronomical Society of the Pacific, 131, 068003,
  \dodoi{10.1088/1538-3873/ab0c2a}

\bibitem[{Bhatti {et~al.}(2020)Bhatti, Bouma, Joshua, John, \&
  Price-Whelan}]{BhBo2020}
Bhatti, W., Bouma, L., Joshua, John, \& Price-Whelan, A. 2020,
  \dodoi{10.5281/zenodo.3723832}

\bibitem[{Breivik {et~al.}(2020)Breivik, Mingarelli, \& Larson}]{BrMi2019}
Breivik, K., Mingarelli, C. M.~F., \& Larson, S.~L. 2020, Constraining Galactic
  Structure with the {LISA} White Dwarf Foreground,  American Astronomical
  Society, \dodoi{10.3847/1538-4357/abab99}

\bibitem[{Burdge {et~al.}(2019{\natexlab{a}})Burdge, Coughlin, Fuller,
  {et~al.}}]{BuCo2019}
Burdge, K.~B., Coughlin, M.~W., Fuller, J., {et~al.} 2019{\natexlab{a}},
  Nature, 571, 528, \dodoi{10.1038/s41586-019-1403-0}

\bibitem[{Burdge {et~al.}(2019{\natexlab{b}})Burdge, Fuller, Phinney, van
  Roestel, Claret, Cukanovaite, Fusillo, Coughlin, Kaplan, Kupfer, Tremblay,
  Dekany, Duev, Feeney, Riddle, Kulkarni, \& Prince}]{BuFu2019}
Burdge, K.~B., Fuller, J., Phinney, E.~S., {et~al.} 2019{\natexlab{b}}, The
  Astrophysical Journal, 886, L12, \dodoi{10.3847/2041-8213/ab53e5}

\bibitem[{Cannizzo \& Nelemans(2015)}]{CaNe2015}
Cannizzo, J.~K., \& Nelemans, G. 2015, The Astrophysical Journal, 803, 19,
  \dodoi{10.1088/0004-637x/803/1/19}

\bibitem[{{Catelan}(2009)}]{Ca2009}
{Catelan}, M. 2009, \apss, 320, 261, \dodoi{10.1007/s10509-009-9987-8}

\bibitem[{Chen \& Guestrin(2016)}]{ChGu2016}
Chen, T., \& Guestrin, C. 2016, Proceedings of the 22nd ACM SIGKDD
  International Conference on Knowledge Discovery and Data Mining,
  \dodoi{10.1145/2939672.2939785}

\bibitem[{Coughlin {et~al.}(2020)Coughlin, Burdge, Sterl Phinney, van Roestel,
  Bellm, Dekany, Delacroix, Duev, Feeney, Graham, \& et~al.}]{CoBu2020}
Coughlin, M.~W., Burdge, K., Sterl Phinney, E., {et~al.} 2020, Monthly Notices
  of the Royal Astronomical Society: Letters, 494, L91–L96,
  \dodoi{10.1093/mnrasl/slaa044}

\bibitem[{{Coughlin et al.}(2019{\natexlab{a}})}]{CoAh2019}
{Coughlin et al.} 2019{\natexlab{a}}, Publications of the Astronomical Society
  of the Pacific, 131, 048001, \dodoi{10.1088/1538-3873/aaff99}

\bibitem[{{Coughlin et al.}(2019{\natexlab{b}})}]{CoAh2019b}
---. 2019{\natexlab{b}}, The Astrophysical Journal, 885, L19,
  \dodoi{10.3847/2041-8213/ab4ad8}

\bibitem[{Dekany {et~al.}(2020)Dekany, Smith, Riddle, Feeney, Porter, Hale,
  Zolkower, Belicki, Kaye, Henning, Walters, Cromer, Delacroix, Rodriguez,
  Reiley, Mao, Hover, Murphy, Burruss, Baker, Kowalski, Reif, Mueller, Bellm,
  Graham, \& Kulkarni}]{DeSm2018}
Dekany, R., Smith, R.~M., Riddle, R., {et~al.} 2020, Publications of the
  Astronomical Society of the Pacific, 132, 038001,
  \dodoi{10.1088/1538-3873/ab4ca2}

\bibitem[{Drake {et~al.}(2009)Drake, Djorgovski, Mahabal, Beshore, Larson,
  Graham, Williams, Christensen, Catelan, Boattini, \& et~al.}]{DrDj2009}
Drake, A.~J., Djorgovski, S.~G., Mahabal, A., {et~al.} 2009, The Astrophysical
  Journal, 696, 870–884, \dodoi{10.1088/0004-637x/696/1/870}

\bibitem[{Drake {et~al.}(2014)Drake, Graham, Djorgovski, Catelan, Mahabal,
  Torrealba, García-Álvarez, Donalek, Prieto, Williams, \& et~al.}]{DrGr2014}
Drake, A.~J., Graham, M.~J., Djorgovski, S.~G., {et~al.} 2014, The
  Astrophysical Journal Supplement Series, 213, 9,
  \dodoi{10.1088/0067-0049/213/1/9}

\bibitem[{Duev {et~al.}(2019)Duev, Mahabal, Masci, Graham, Rusholme, Walters,
  Karmarkar, Frederick, Kasliwal, Rebbapragada, \& Ward}]{DuMa2019}
Duev, D.~A., Mahabal, A., Masci, F.~J., {et~al.} 2019, Monthly Notices of the
  Royal Astronomical Society, 489, 3582, \dodoi{10.1093/mnras/stz2357}

\bibitem[{{Freedman} {et~al.}(2001){Freedman}, {Madore}, {Gibson}, {Ferrarese},
  {Kelson}, {Sakai}, {Mould}, {Kennicutt}, {Ford}, {Graham}, {Huchra},
  {Hughes}, {Illingworth}, {Macri}, \& {Stetson}}]{FrMa2001}
{Freedman}, W.~L., {Madore}, B.~F., {Gibson}, B.~K., {et~al.} 2001, \apj, 553,
  47, \dodoi{10.1086/320638}

\bibitem[{Fuller \& Lai(2011)}]{FuLa2011}
Fuller, J., \& Lai, D. 2011, Monthly Notices of the Royal Astronomical Society,
  412, 1331, \dodoi{10.1111/j.1365-2966.2010.18017.x}

\bibitem[{{Gaia Collaboration}(2018)}]{Gaia2018}
{Gaia Collaboration}. 2018, A\&A, 616, A1, \dodoi{10.1051/0004-6361/201833051}

\bibitem[{Graham {et~al.}(2013{\natexlab{a}})Graham, Drake, Djorgovski,
  Mahabal, \& Donalek}]{GrDr2013}
Graham, M.~J., Drake, A.~J., Djorgovski, S.~G., Mahabal, A.~A., \& Donalek, C.
  2013{\natexlab{a}}, Monthly Notices of the Royal Astronomical Society, 434,
  2629, \dodoi{10.1093/mnras/stt1206}

\bibitem[{Graham {et~al.}(2013{\natexlab{b}})Graham, Drake, Djorgovski,
  Mahabal, Donalek, Duan, \& Maker}]{GrDr2013b}
Graham, M.~J., Drake, A.~J., Djorgovski, S.~G., {et~al.} 2013{\natexlab{b}},
  Monthly Notices of the Royal Astronomical Society, 434, 3423–3444,
  \dodoi{10.1093/mnras/stt1264}

\bibitem[{{Graham et al.}(2019)}]{Graham2018}
{Graham et al.} 2019, Publications of the Astronomical Society of the Pacific,
  131, 078001, \dodoi{10.1088/1538-3873/ab006c}

\bibitem[{{Hoffleit} \& {Jaschek}(1991)}]{HoJa1991}
{Hoffleit}, D., \& {Jaschek}, C. 1991, {The Bright star catalogue}

\bibitem[{Huber {et~al.}(2006)Huber, Everett, \& Howell}]{HuEv2006}
Huber, M.~E., Everett, M.~E., \& Howell, S.~B. 2006, The Astronomical Journal,
  132, 633–649, \dodoi{10.1086/505300}

\bibitem[{Ivezic {et~al.}(2019)Ivezic, Tyson, Allsman, Andrew, \&
  Angel}]{Ivezic2014}
Ivezic, Z., Tyson, J.~A., Allsman, R., Andrew, J., \& Angel, R. 2019, \apj,
  873, 111, \dodoi{10.3847/1538-4357/ab042c}

\bibitem[{Katz {et~al.}(2020)Katz, Cooper, Coughlin, Burdge, Breivik, \&
  Larson}]{KaCo2020}
Katz, M.~L., Cooper, O.~R., Coughlin, M.~W., {et~al.} 2020, GPU-Accelerated
  Periodic Source Identification in Large-Scale Surveys: Measuring $P$ and
  $\dot{P}$.
\newblock \doarXiv{2006.06866}

\bibitem[{Kochanek {et~al.}(2017)Kochanek, Shappee, Stanek, Holoien, Thompson,
  Prieto, Dong, Shields, Will, Britt, Perzanowski, \&
  Pojma{\'{n}}ski}]{KoSh2017}
Kochanek, C.~S., Shappee, B.~J., Stanek, K.~Z., {et~al.} 2017, Publications of
  the Astronomical Society of the Pacific, 129, 104502,
  \dodoi{10.1088/1538-3873/aa80d9}

\bibitem[{Kremer {et~al.}(2018)Kremer, Chatterjee, Breivik, Rodriguez, Larson,
  \& Rasio}]{KrCh2018}
Kremer, K., Chatterjee, S., Breivik, K., {et~al.} 2018, Physical Review
  Letters, 120, \dodoi{10.1103/physrevlett.120.191103}

\bibitem[{Kupfer {et~al.}(2018)Kupfer, Korol, Shah, Nelemans, Marsh, Ramsay,
  Groot, Steeghs, \& Rossi}]{KuKo2018}
Kupfer, T., Korol, V., Shah, S., {et~al.} 2018, Monthly Notices of the Royal
  Astronomical Society, 480, 302–309, \dodoi{10.1093/mnras/sty1545}

\bibitem[{{Kupfer} {et~al.}(2019){Kupfer}, {Bauer}, {Burdge}, {Bellm},
  {Bildsten}, {Fuller}, {Hermes}, {Kulkarni}, {Prince}, {van Roestel},
  {Dekany}, {Duev}, {Feeney}, {Giomi}, {Graham}, {Kaye}, {Laher}, {Masci},
  {Porter}, {Riddle}, {Shupe}, {Smith}, {Soumagnac}, {Szkody}, \&
  {Ward}}]{KuBa2019}
{Kupfer}, T., {Bauer}, E.~B., {Burdge}, K.~B., {et~al.} 2019, \apjl, 878, L35,
  \dodoi{10.3847/2041-8213/ab263c}

\bibitem[{Kupfer {et~al.}(2020)Kupfer, Bauer, Burdge, van Roestel, Bellm,
  Fuller, Hermes, Marsh, Bildsten, Kulkarni, Phinney, Prince, Szkody, Yao,
  Irrgang, Heber, Schneider, Dhillon, Murawski, Drake, Duev, Feeney, Graham,
  Laher, Littlefair, Mahabal, Masci, Porter, Reiley, Rodriguez, Rusholme,
  Shupe, \& Soumagnac}]{KuBa2020}
Kupfer, T., Bauer, E.~B., Burdge, K.~B., {et~al.} 2020, The Astrophysical
  Journal, 898, L25, \dodoi{10.3847/2041-8213/aba3c2}

\bibitem[{{Lomb}(1976)}]{Lo1976}
{Lomb}, N.~R. 1976, \apss, 39, 447, \dodoi{10.1007/BF00648343}

\bibitem[{Lopes {et~al.}(2020)Lopes, Cross, Catelan, Minniti, Hempel, Lucas,
  Angeloni, Jablonsky, Braga, Leao, Herpich, Alonso-Garcia, Papageorgiou,
  Pichara, Saito, Bradley, Beamin, Cortes, Medeiros, \& Russell}]{LoCr2020}
Lopes, C. E.~F., Cross, N. J.~G., Catelan, M., {et~al.} 2020, The VVV Infrared
  Variability Catalog (VIVA-I).
\newblock \doarXiv{2005.05404}

\bibitem[{Masci {et~al.}(2018)Masci, Laher, Rusholme, Shupe, Groom, Surace,
  Jackson, Monkewitz, Beck, Flynn, \& et~al.}]{MaLa2018}
Masci, F.~J., Laher, R.~R., Rusholme, B., {et~al.} 2018, Publications of the
  Astronomical Society of the Pacific, 131, 018003,
  \dodoi{10.1088/1538-3873/aae8ac}

\bibitem[{McGibbon \& Zhao(2019)}]{CUDAwrapper}
McGibbon, R., \& Zhao, Y. 2019, npcuda-example.
\newblock \url{https://github.com/rmcgibbo/npcuda-example}

\bibitem[{Morgan {et~al.}(2012)Morgan, Kaiser, Moreau, Anderson, \&
  Burgett}]{MoKa2012}
Morgan, J.~S., Kaiser, N., Moreau, V., Anderson, D., \& Burgett, W. 2012, Proc.
  SPIE Int. Soc. Opt. Eng., 8444, 0H, \dodoi{10.1117/12.926646}

\bibitem[{{Mortier, A.} \& {Collier Cameron, A.}(2017)}]{MoCo2017}
{Mortier, A.}, \& {Collier Cameron, A.} 2017, A\&A, 601, A110,
  \dodoi{10.1051/0004-6361/201630201}

\bibitem[{{Mortier, A.} {et~al.}(2015){Mortier, A.}, {Faria, J. P.}, {Correia,
  C. M.}, {Santerne, A.}, \& {Santos, N. C.}}]{MoFa2015}
{Mortier, A.}, {Faria, J. P.}, {Correia, C. M.}, {Santerne, A.}, \& {Santos, N.
  C.} 2015, A\&A, 573, A101, \dodoi{10.1051/0004-6361/201424908}

\bibitem[{Naul {et~al.}(2016)Naul, van~der Walt, Crellin-Quick, Bloom, \&
  Pérez}]{NaWa2016}
Naul, B., van~der Walt, S., Crellin-Quick, A., Bloom, J.~S., \& Pérez, F.
  2016, cesium: Open-Source Platform for Time-Series Inference.
\newblock \doarXiv{1609.04504}

\bibitem[{Nelemans \& Tout(2005)}]{NeTo2005}
Nelemans, G., \& Tout, C.~A. 2005, Monthly Notices of the Royal Astronomical
  Society, 356, 753, \dodoi{10.1111/j.1365-2966.2004.08496.x}

\bibitem[{{Ngeow} {et~al.}(2019){Ngeow}, {Lee}, {Yu}, {Masci}, {Laher},
  {Kupfer}, {Golkhou}, \& {ZTF Collaboration}}]{NgLe2019}
{Ngeow}, C.~C., {Lee}, C.~D., {Yu}, P.~C., {et~al.} 2019, in Journal of Physics
  Conference Series, Vol. 1231, Journal of Physics Conference Series, 012010,
  \dodoi{10.1088/1742-6596/1231/1/012010}

\bibitem[{Nickolls {et~al.}(2008)Nickolls, Buck, Garland, \& Skadron}]{CUDA}
Nickolls, J., Buck, I., Garland, M., \& Skadron, K. 2008, Queue, 6, 40,
  \dodoi{10.1145/1365490.1365500}

\bibitem[{Nun {et~al.}(2015)Nun, Protopapas, Sim, Zhu, Dave, Castro, \&
  Pichara}]{NuPr2015}
Nun, I., Protopapas, P., Sim, B., {et~al.} 2015, FATS: Feature Analysis for
  Time Series.
\newblock \doarXiv{1506.00010}

\bibitem[{Pashchenko {et~al.}(2017)Pashchenko, Sokolovsky, \&
  Gavras}]{PaSo2017}
Pashchenko, I.~N., Sokolovsky, K.~V., \& Gavras, P. 2017, Monthly Notices of
  the Royal Astronomical Society, 475, 2326–2343,
  \dodoi{10.1093/mnras/stx3222}

\bibitem[{{Saha}(1984)}]{Sa1984}
{Saha}, A. 1984, \apj, 283, 580, \dodoi{10.1086/162343}

\bibitem[{{Saha}(1985)}]{Sa1985}
---. 1985, \apj, 289, 310, \dodoi{10.1086/162890}

\bibitem[{{Scargle}(1982)}]{Sc1982}
{Scargle}, J.~D. 1982, \apj, 263, 835, \dodoi{10.1086/160554}

\bibitem[{{Schwarzenberg-Czerny}(1998)}]{ScCz1998}
{Schwarzenberg-Czerny}, A. 1998, Baltic Astronomy, 7, 43,
  \dodoi{10.1515/astro-1998-0109}

\bibitem[{Shapiro \& Wilk(1965)}]{ShWi1965}
Shapiro, S.~S., \& Wilk, M.~B. 1965, Biometrika, 52, 591,
  \dodoi{10.1093/biomet/52.3-4.591}

\bibitem[{Shappee {et~al.}(2014)Shappee, Prieto, Grupe, Kochanek, Stanek, Rosa,
  Mathur, Zu, Peterson, Pogge, Komossa, Im, Jencson, Holoien, Basu, Beacom,
  Szczygie{\l}, Brimacombe, Adams, Campillay, Choi, Contreras, Dietrich,
  Dubberley, Elphick, Foale, Giustini, Gonzalez, Hawkins, Howell, Hsiao, Koss,
  Leighly, Morrell, Mudd, Mullins, Nugent, Parrent, Phillips, Pojmanski,
  Rosing, Ross, Sand, Terndrup, Valenti, Walker, \& Yoon}]{ShPr2014}
Shappee, B.~J., Prieto, J.~L., Grupe, D., {et~al.} 2014, The Astrophysical
  Journal, 788, 48, \dodoi{10.1088/0004-637x/788/1/48}

\bibitem[{Stephens(1974)}]{St1974}
Stephens, M.~A. 1974, Journal of the American Statistical Association, 69, 730.
\newblock \url{http://www.jstor.org/stable/2286009}

\bibitem[{{Stetson}(1996)}]{St1996}
{Stetson}, P.~B. 1996, \pasp, 108, 851, \dodoi{10.1086/133808}

\bibitem[{Szkody {et~al.}(2020)Szkody, Dicenzo, Ho, Hillenbrand, van Roestel,
  Ridder, DeJesus~Lima, Graham, Bellm, Burdge, \& et~al.}]{SzDi2020}
Szkody, P., Dicenzo, B., Ho, A. Y.~Q., {et~al.} 2020, The Astronomical Journal,
  159, 198, \dodoi{10.3847/1538-3881/ab7cce}

\bibitem[{{Tiwari} {et~al.}(2015){Tiwari}, {Gupta}, {Rogers}, {Maxwell},
  {Rech}, {Vazhkudai}, {Oliveira}, {Londo}, {DeBardeleben}, {Navaux}, {Carro},
  \& {Bland}}]{TiGu2015}
{Tiwari}, D., {Gupta}, S., {Rogers}, J., {et~al.} 2015, in 2015 IEEE 21st
  International Symposium on High Performance Computer Architecture (HPCA),
  331--342

\bibitem[{Tonry {et~al.}(2018)Tonry, Denneau, Heinze, Stalder, Smith, Smartt,
  Stubbs, Weiland, \& Rest}]{ToDe2018}
Tonry, J.~L., Denneau, L., Heinze, A.~N., {et~al.} 2018, Publications of the
  Astronomical Society of the Pacific, 130, 064505.
\newblock \url{http://stacks.iop.org/1538-3873/130/i=988/a=064505}

\bibitem[{Towns {et~al.}(2014)Towns, Cockerill, Dahan, Foster, Gaither,
  Grimshaw, Hazlewood, Lathrop, Lifka, Peterson, Roskies, Scott, \&
  Wilkins-Diehr}]{ToCo2014}
Towns, J., Cockerill, T., Dahan, M., {et~al.} 2014, Computing in Science \&
  Engineering, 16, 62, \dodoi{10.1109/MCSE.2014.80}

\bibitem[{{Udalski}(2003)}]{udalski2003}
{Udalski}, A. 2003, \actaa, 53, 291.
\newblock \doarXiv{astro-ph/0401123}

\bibitem[{{Udalski} {et~al.}(2015){Udalski}, {Szyma{\'n}ski}, \&
  {Szyma{\'n}ski}}]{udalski2015}
{Udalski}, A., {Szyma{\'n}ski}, M.~K., \& {Szyma{\'n}ski}, G. 2015, \actaa, 65,
  1.
\newblock \doarXiv{1504.05966}

\bibitem[{{Zechmeister, M.} \& {K\"urster, M.}(2009)}]{ZeKu2009}
{Zechmeister, M.}, \& {K\"urster, M.} 2009, A\&A, 496, 577,
  \dodoi{10.1051/0004-6361:200811296}

\end{thebibliography}

\end{document}